\documentclass[fleqn,10pt]{wlscirep}
\usepackage[utf8]{inputenc}
\usepackage[T1]{fontenc}
\usepackage{amsmath}
\usepackage{graphicx}
\usepackage{dcolumn}
\usepackage{bm}
\usepackage{xcolor}
\usepackage{soul}
\usepackage{gensymb}

\newcommand{\pN}{\ \mathrm{pN}}
\newcommand{\nm}{\ \mathrm{nm}}
\newcommand{\Mm}{\ \mathrm{\mu m}}
\newcommand{\bp}{\ \mathrm{bp}}

\newcommand{\persec}{\ \mathrm{s}^{-1}}

\newcommand{\mMol}{\mathrm{mM}}

\newcommand{\vR}{R}
\newcommand{\vL}{L}
\newcommand{\DR}{\Delta {R}}
\newcommand{\Dl}{\Delta {L}}
\newcommand{\angstrom}{\mbox{\normalfont\AA}}
\title{Theory and Simulations of condensin mediated loop extrusion in DNA}

\author[1]{Ryota Takaki}
\author[2]{Atreya Dey}
\author[2]{Guang Shi}
\author[2*]{D.Thirumalai}
\affil[1]{Department of Physics, The university of Texas at Austin, Austin, 78712, USA}
\affil[2]{Department of Chemistry, The university of Texas at Austin, chemistry, Austin, 78712, USA}

\affil[*]{dave.thirumalai@gmail.com}

\begin{abstract}
Condensation of hundreds of mega-base-pair-long human chromosomes in a small nuclear volume is a spectacular biological phenomenon. This process is  driven  by the formation of chromosome loops. The ATP consuming motor, condensin, interacts with chromatin segments to actively extrude loops. Motivated by real-time imaging of loop extrusion (LE), we created an analytically solvable  model, predicting the LE velocity and step size distribution as a function of external load. The theory fits the available experimental data quantitatively, and suggests that condensin must undergo a large conformational change, induced by ATP binding,  bringing distant parts of the motor to proximity. Simulations using a simple model confirm that the motor transitions between an  open and a closed state in order to extrude loops by a scrunching mechanism, similar to that proposed in DNA bubble formation during bacterial transcription. Changes in the orientation of the motor domains are transmitted over $\sim$ $50\nm$, connecting the motor head and the hinge, thus providing an allosteric basis for LE.
\end{abstract}
\begin{document}

\flushbottom
\maketitle
%
%
\thispagestyle{empty}


\section*{Introduction}
How chromosomes are structurally organized in the tight space of the nucleus  is a long-standing problem in biology. Remarkably, these information-carrying polymers in humans with more than 100 million base pairs, are densely packed in the $5-10 \Mm$ cell nucleus~\cite{flemming1882zellsubstanz,alberts2013essential}. In order to accomplish this herculean feat, nature has evolved a family of SMC (Structural Maintenance of Chromosomes) complexes~\cite{hagstrom2003condensin,yatskevich2019organization} (bacterial SMC, cohesin, and condensin) to enable large scale compaction of chromosomes in both prokaryotic and eukaryotic systems. Compaction is thought to occur by an active generation of a large array of loops, which are envisioned to form by extrusion of the genomic material~\cite{nasmyth2001disseminating,fudenberg2016formation,sanborn2015chromatin} driven by ATP-consuming motors. The SMC complexes have been identified as a major component of the loop extrusion (LE) process~\cite{hagstrom2003condensin,yatskevich2019organization}. 

Of interest here is condensin, whose motor activity~\cite{terakawa2017condensin}, results in  active extrusion of loops in an ATP-dependent manner~\cite{ganji2018real}. Let us first describe the architecture of condensin, shown schematically in Fig~\ref{fig:modelfig}. Condensin is a ring-shaped dimeric motor, containing a pair of  SMC proteins (Smc2 and Smc4).  Both Smc2 and Smc4,  which have coiled-coil (CC) structures, are connected at the hinge domain. The  ATP binding domains are in the motor heads~\cite{diebold2017structure,yatskevich2019organization}.  There are kinks roughly in the middle of the CCs~\cite{diebold2017structure}. The relative flexibility in the elbow region (located near the kinks) could be the key to the conformational transitions in the CC that are powered by ATP binding and hydrolysis~\cite{buermann2019folded,yatskevich2019organization}. 

Previous studies using simulations~\cite{fudenberg2016formation,alipour2012self,goloborodko2016compaction}, which were built on the pioneering insights by Nasmyth~\cite{nasmyth2001disseminating}, suggested that multiple condensins  translocate along the chromosome extruding loops of increasing length.  In this mechanism, the two condensin heads move away from each other extruding loops in a symmetric manner. Cooperative action of many condensins~\cite{kim2020dna} might be necessary to account for the $\sim (1,000 - 10,000)$ fold compaction of human chromosomes~\cite{banigan2019limits}. The only other theoretical study that predicts LE velocity as a function of an external load~\cite{marko2019dna} is based on a four-state stochastic kinetic model, with minimally twenty parameters, for the catalytic cycle of the condensin that is coupled to loop extrusion~\cite{marko2019dna}. 

In sharp contrast, by focusing on the motor activity of condensin through ATP-driven allosteric changes in the enzyme, our theory and simulations support   "scrunching" as a plausible mechanism for  loop extrusion.  Scrunching  is reminiscent of the proposal made over a decade ago in the context of the first stage in bacterial transcription that results  in bubble formation in promoter DNA~\cite{kapanidis2006initial}, which was quantitatively affirmed using molecular simulations~\cite{chen2010promoter}. Recently, the scrunching mechanism was proposed to explain loop extrusion~\cite{ryu2020condensin}, which is fully supported by  theory and simulations presented here.

We were inspired by the real-time imaging of LE in $\lambda$-DNA by a  single condensin~\cite{ganji2018real},  which functions by extruding loops asymmetrically. To describe the experimental outcomes quantitatively, we created a simple analytically solvable theory, with two parameters, that produces excellent agreement with experiments for the LE velocity as a function of external load. We also quantitatively reproduce the distribution of LE length per cycle measured using magnetic tweezer experiments~\cite{ryu2020resolving}. The theory and simulations show that  for LE to occur there has to be an ATP-powered  allosteric transition in condensin  involving a large  conformational change that brings distant parts (head and the hinge in Fig~\ref{fig:modelfig}) of the motor to proximity. We predict that, on an average,  the distance between the head and the hinge decreases by conformational $\sim (22- 26)\nm$ per catalytic cycle. These values are in remarkably close to  the experimentally inferred values~\cite{ganji2018real}. Simulations using a simple model, with and without DNA, lend support to our findings. Our work strongly suggests that the conformational transitions are driven by a scrunching mechanism in which the motor is relatively stationary but DNA is reeled in by an allosteric mechanism.  


\section*{Results}


\subsection*{Model description}
\begin{figure}[]
\centering
\includegraphics[width=0.5\textwidth]{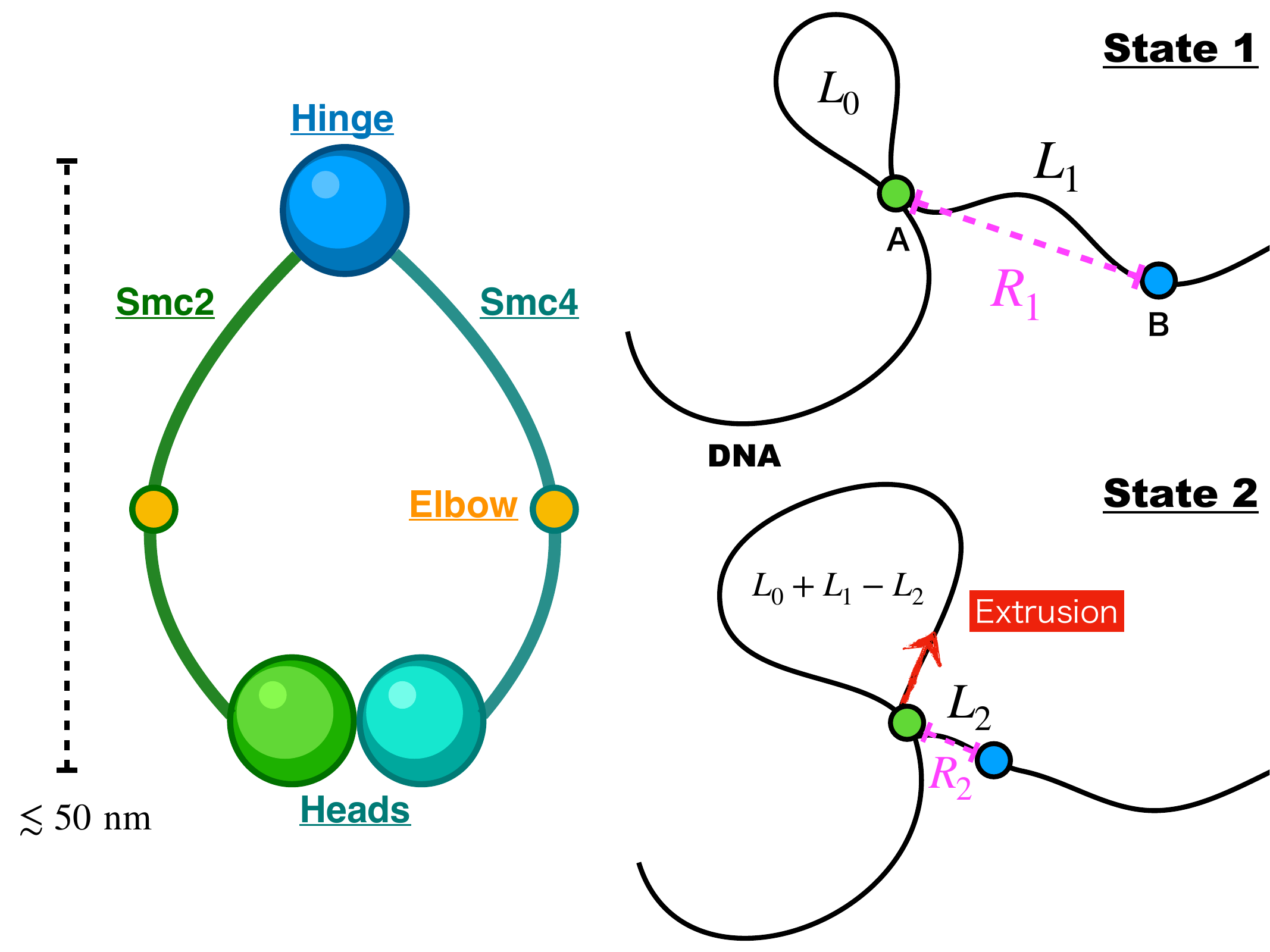}
\caption{\label{fig:modelfig} {\bf Left panel:} Caricature of the structure of condensin, which has two heads (ATPase domains) and a hinge connected by coiled-coils, labeled Smc2 and Smc4. In the middle of the CCs, there is a flexible kink, referred to as an elbow. {\bf Right panel:} A schematic of the physical picture for one-sided loop extrusion based on the architecture of a generic SMC complex. DNA is attached to two structural regions on condensin. In state 1 (upper panel) the conformation of condensin is extended with the spatial distance between {\bf A} and {\bf B} equal to  $R_1$. The genomic length at the  start is $L_0$, which can be large or small.  After the conformational transition (state 1 to state 2) the distance between {\bf A} and {\bf B} shrinks to  $R_2$, and the length of the extrusion during the single transition is $\Dl =L_1 - L_2$, which would vary from cycle to cycle. }
\end{figure}

In order to develop a model applicable to condensin (and cohesin), we assume that condensin is attached to  two loci ({\bf A} and {\bf B}) on the DNA (Fig.\ref{fig:modelfig}; right panel). Although we do not explicitly describe the nature of the attachment points, our model is based on the idea of scrunching motion where two distant ends of condensin move closer upon conformational change, triggered by ATP binding. For example, the green and blue sphere may be mapped onto motor heads and hinge, respectively.   
The structure of condensin-DNA complex in the LE active form is currently unavailable. However,  cryo-EM structures for the related cohesin-DNA complex~\cite{shi2020cryo} reveal that DNA is tightly gripped by the two heads of cohesin and the subunits (NIPBL and RAD21). When the results of structural studies are integrated with the observation that the hinge domain of the SMC complexes binds to DNA~\cite{chiu2004dna,griese2010structure,alt2017specialized}, we conclude that both condensin and cohesin must use a similar mechanism to engage with  DNA. The head domains in these motors interact with the DNA segment that is in proximity whereas DNA binds only  transiently to the hinge. We constructed the model in Fig.\ref{fig:modelfig} based in part on these findings.

In state 1, the spatial distance between the condensin attachment points is, $R_1$, and  the genomic length between {\bf A} and {\bf B} is $L_1$. Due to the polymeric nature of the DNA, the captured length $L_1$ could exceed $R_1$. However, $R_1$ cannot be greater than the overall dimension of the SMC motor, which is on the order of $\sim 50\nm$. Once a segment in the DNA is captured, condensin undergoes a conformational change driven most likely by ATP binding~\cite{ryu2020resolving}, shrinking the distance from $R_1$ to $R_2$ (where $R_2 < R_1$). As a result, the captured genomic length between {\bf A} and {\bf B}  reduces  to $L_2$ (state 2). Consequently, the loop grows by $L_1-L_2$ . The step size of condensin is $\DR=R_1-R_2$, and extrusion length per step is $\Dl=L_1-L_2$. After the extrusion is completed one end of condensin (blue circle in Fig.\ref{fig:modelfig}; right panel) is released from the genome segment and starts the DNA capturing process again, likely mediated by diffusion leading to the next LE cycle.


\subsection*{Theory for the captured length ($\vL$) of DNA}
\begin{figure}[]
\centering
\includegraphics[width=0.5\textwidth]{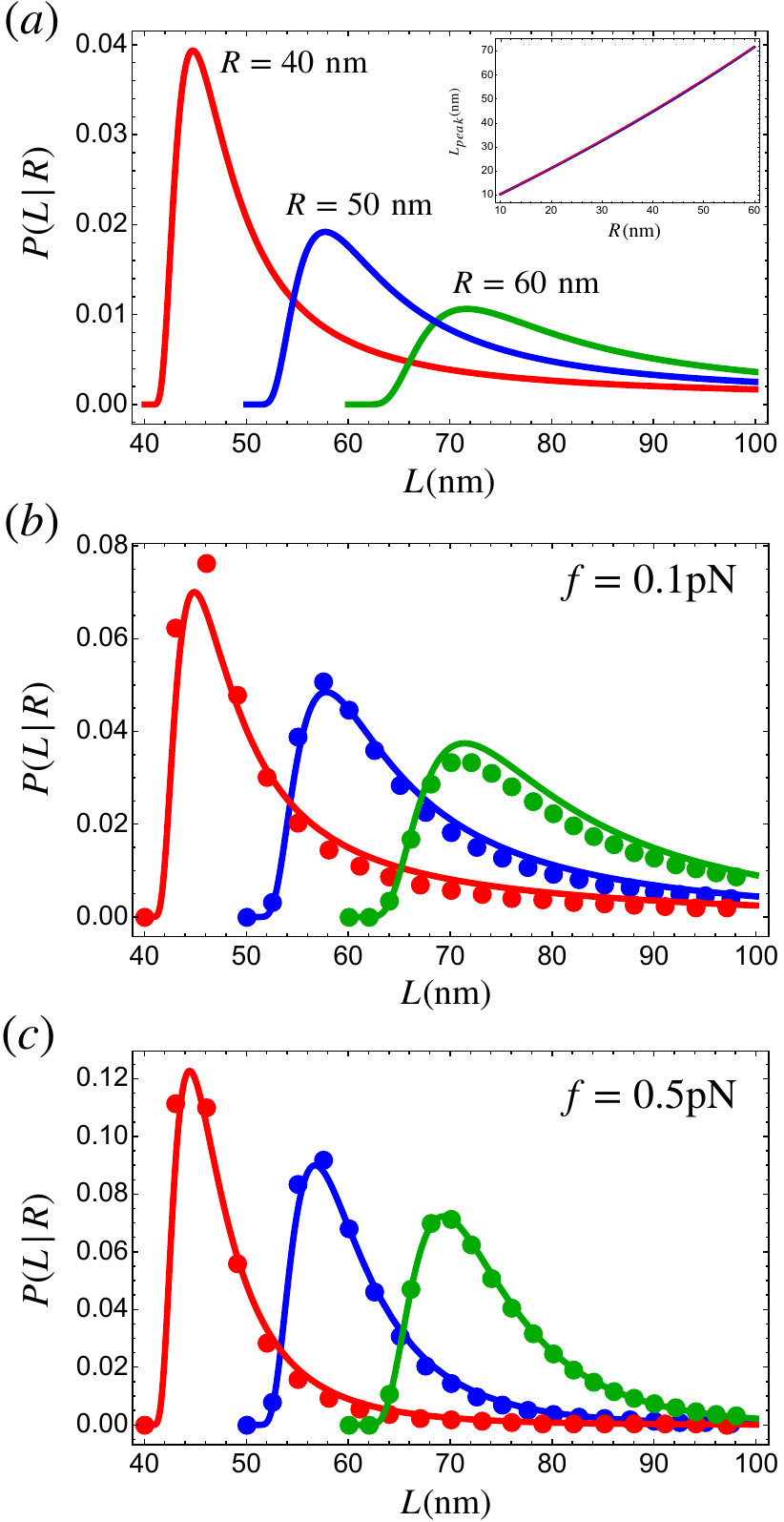}
\caption{\label{fig:PLFdist} (a) Plots of $P(\vL|\vR)$ for different $\vR$ values; $\vR=40\nm$ (red), $\vR=50\nm$ (blue), and $\vR=60\nm$ (green). Inset: Peak position ($\vL_{peak}$) of $P(\vL|\vR)$, evaluated numerically by setting by $dP(\vL|\vR)/d\vL$ to zero, as a function of $\vR$. The dotted red line is a fit, $\vL_{peak}=\vR\exp(a\vR)$ with $a=0.003\ \mathrm{nm}^{-1}$.   
($b$)-($c$): The distributions of $L$ for different $\vR$ and $f$. $\vR=40\nm$ (red), $\vR=50\nm$ (blue), and $\vR=60\nm$ (green). The dots are from Eq.(\ref{eq:P(L|R,f)}) and the solid lines are the approximate probability distribution Eq.(\ref{eq:P(L|R,f)app}).  We used $l^{DNA}_p=50\nm$.
}
\end{figure}

In order to derive an expression for the loop extrusion velocity, we first estimate the loop length of DNA, $\vL$, captured by condensin when the attachment points are spatially separated by $\vR$. 
We show that on the length scale of the size of condensin ($\sim 50\nm$), it is reasonable to approximate $\vL \approx \vR $. To calculate the LE velocity it is necessary to estimate the total work done to extrude DNA with and without an external load. Based on these considerations, we derive an expression for the LE velocity, given by $k_0 \exp({-f\DR/k_BT})\DR$, where $k_0$ is the rate of mechanical step in the absence of the external load ($f$), $k_B$ is Boltzmann constant and $T$ is temperature.

\subsubsection*{Distribution of captured DNA length without tension, $\boldsymbol{P(\vL|\vR)}$ }
We examined the possibility that the loop extrusion length per step can be considerably larger than the size of condensin~\cite{terakawa2017condensin,lawrimore2017rotostep,ganji2018real,diebold2017structure} by calculating, $P(\vL|\vR)$, the conditional probability for realizing the contour length $\vL$ for a given end-to-end distance, $\vR$.  We calculated $P(\vR|\vL)$ using a mean-field theory that is an excellent approximation~\cite{hyeon2006kinetics} to the exact but complicated expression~\cite{wilhelm1996radial}.  The expression for $P(\vL|\vR)$, which has the same form as $P(\vR|\vL)$, up to a normalization constant, is given by (Sec.III in SI) 
\begin{align} 
\begin{split}
\label{eq:P(L|R)}
P(\vL|\vR) = A \frac{4\pi N\{\vL \} (\vR/\vL)^2}{\vL(1-(\vR/\vL)^2)^{9/2}} \exp\Big(-\frac{3t\{\vL\}}{4(1-(\vR/\vL)^2)}\Big),
\end{split} 
\end{align}
where $t\{\vL\}=3\vL/2l_p$, $l_p$ is the persistence length of the polymer, and  $N\{\vL\}=\frac{4\alpha^{3/2}e^{\alpha}}{\pi^{3/2}(4+12\alpha^{-1}+15\alpha^{-2})}$ with $\alpha\{\vL\}=3t/4$. In Eq.(\ref{eq:P(L|R)}), $A$ is a normalization constant that does not depend on $\vL$ where integration range for $\vL$ is from $\vR$ to $\infty$. 
The distribution $P(\vL|\vR)$, which scales as $\vL^{-3/2}$ for large $\vL$, has a heavy tail and does not have a well defined mean (see Fig.~\ref{fig:PLFdist}a for the plots of $P(\vL|\vR)$ for different $\vR$). The existence of long-tail in the distribution $P(\vL|\vR)$ already suggests that condensin, in principle, could capture DNA segment much larger than its size $\sim 50\nm$. However, this can only happen with lower probability compared to a more probable scenario where $\vR \approx \vL$ near the position of the peak in $P(\vL|\vR)$ (see Fig.\ref{fig:PLFdist}a). Therefore, we evaluated the location of the peak ($\vL_{peak}$)  in $P(\vL|\vR)$, and solved the resulting equation numerically. The dependence of $\vL_{peak}$ on $\vR$, which is almost linear (Fig.\ref{fig:PLFdist}a),  is well fit using $\vL_{peak}=\vR\exp(a\vR)$ with $a=0.003\ \mathrm{nm}^{-1}$ for $\vR < 60\nm$. Thus, with negligible corrections, we used the approximation $\vL \approx \vR$ on the length scales corresponding to the size of condensin or the DNA persistence length. 
Indeed, the location of the largest probability is at  $\vL\approx l_p \approx \vR$, which is similar to what was found for proteins~\cite{thirumalai1999time} as well. 
The presence of $f$ would stretch the DNA, in turn decrease the length of DNA that condensin captures, further justifying the assumption ($\vL\approx \vR$).  Therefore, we conclude that $R_1 \approx L_1$ and $R_2\approx L_2$ (note that $R_2 < R_1 \lesssim l^{DNA}_p$), and that LE of  DNA loop that is much larger than the size of condensin is less likely. 
Thus, we expect that the  extrusion length of DNA is nearly equal to the step size of condensin, $\DR \approx \Dl$.  




\subsubsection*{Force-dependent distribution, $\boldsymbol{P(\vL|\vR,f)}$, of the captured DNA length}
Following the steps described above, we write the 
end-to-end distribution of semi-flexible polymer under tension $f$, $P(\vR,f|\vL)$, as $P(\vR,f|\vL) = B P(\vR|\vL)e^{f R/k_BT}$, where $B$ is normalization constant. Thus, $P(\vL|\vR,f)$ is obtained using,
\begin{align} 
\begin{split}
\label{eq:P(L|R,f)}
P(\vL|\vR,f) = C B\{\vL \} \frac{4\pi N\{\vL \} (\vR/\vL)^2}{\vL(1-(\vR/\vL)^2)^{9/2}} \exp\Big(-\frac{3t\{\vL\}}{4(1-(\vR/\vL)^2)}\Big) \exp\Big(\frac{f\vR}{k_BT}  \Big)  ,
\end{split} 
\end{align}
where $C$ is a normalization constant that does not depend on $\vL$. The constant $B\{\vL \}$ for $P(\vR,f|\vL)$, which carries the $\vL$ dependence, prevents us from deriving an analytically tractable expression for $P(\vL|\vR,f)$. We find that, for a sufficiently stiff polymer, ($\vR/l_p \lesssim 1$), $P(\vL|\vR,f)$ is well approximated by,
\begin{align} 
\begin{split}
\label{eq:P(L|R,f)app}
P_{A}(\vL|\vR,f >  0) = D \frac{4\pi N^2\{\vL \} (\vR/\vL)^2}{\vL(1-(\vR/\vL)^2)^{9/2}} \exp\Big(-\frac{3t\{\vL\}}{4(1-(\vR/\vL)^2)}\Big) \exp\Big(\frac{f\vR}{k_BT} -\big(1.0+3.3e^{-f/f_0}\big) \frac{f \vL}{k_BT} \Big) ,
\end{split} 
\end{align}
where $f_0=1/7\pN$, $t\{\vL\}=3\vL/2l_p$, $l_p$ is the persistence length of the polymer, and  $N^2\{\vL\}=\Big(\frac{4\alpha^{3/2}e^{\alpha}}{\pi^{3/2}(4+12\alpha^{-1}+15\alpha^{-2})}\Big)^2$ with $\alpha\{\vL\}=3t/4$. The constant $D$  does not depend on $\vL$ in the integration range. Note that on the scale of condensin DNA is relatively rigid ($R \lesssim l^{DNA}_p \sim 50\nm$). Probability Eq.(\ref{eq:P(L|R,f)}) and Eq.(\ref{eq:P(L|R,f)app}) for different values of $f$ in Fig.\ref{fig:PLFdist} show that the agreement between $P_{A}(\vL|\vR)$ and $P(\vL|\vR,f)$ is good. Therefore, in what follows we use  Eq.(\ref{eq:P(L|R,f)app}). The distributions for $f>0$ in Fig.\ref{fig:PLFdist} show that the position of the peak for $ P(\vL|\vR,f)$ do not change over the range of $f$ of interest. However, the height of the peak increases as $f$ increases accompanied by the shrinking of the tail for large $\vL$.


\subsection*{Condensin converts chemical energy into mechanical work for LE}

\begin{figure}[]
\centering
\includegraphics[width=0.5\textwidth]{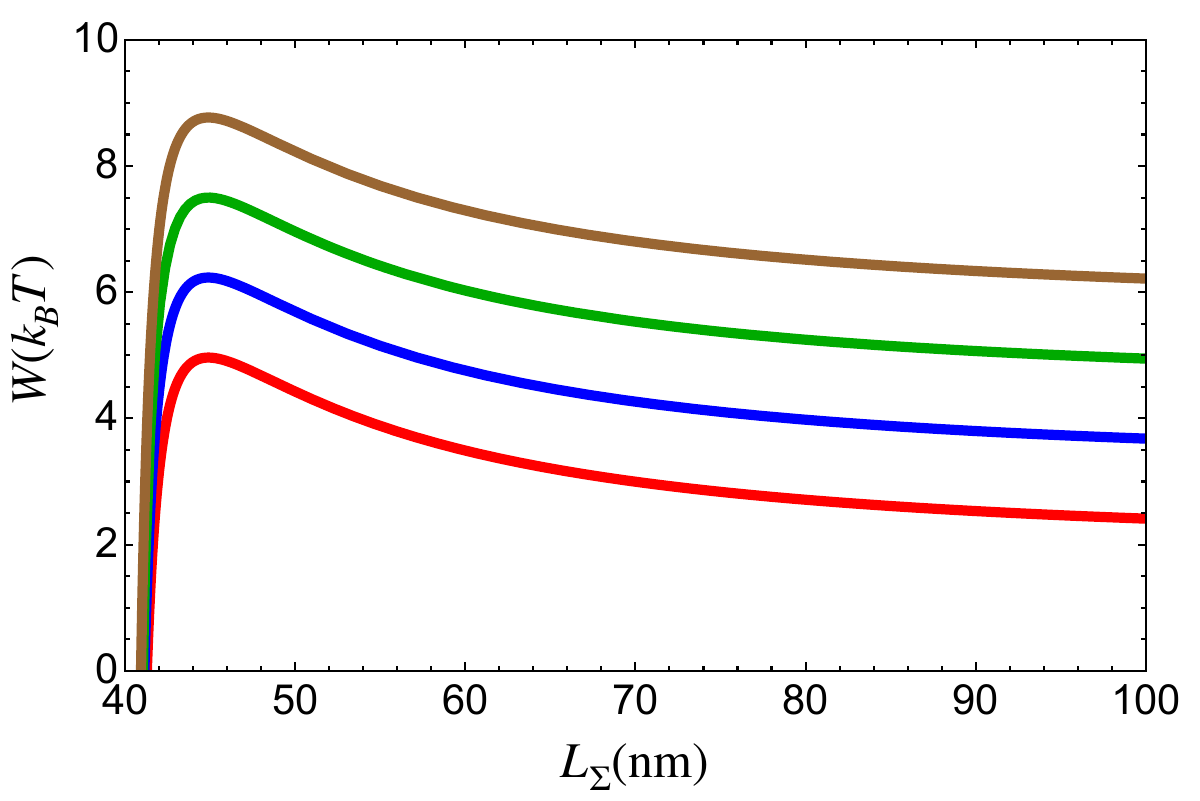}
\caption{\label{fig:delF} Energetic cost to extrude DNA, $W=W_0+W_f$, as a function of $L_\Sigma$ ($\sim$ total extruded length of DNA; $L_{\Sigma}=L_0+L_1$) for different $f$ values on DNA. 
We set $R_1 = 40 \nm$ as the size of condensin in open state (State 1 in Fig.\ref{fig:modelfig}) and $R_2=14\nm$, which corresponds to $\Delta R=26 \nm$ as obtained from the fit of our theory to experiment~\cite{ganji2018real} in later section. $f=0\pN$ is in red, $f=0.2\pN$ is in blue, $f=0.4\pN$ is in green, and $f=0.6\pN$ is in brown.  We used $l^{DNA}_p=50\nm$.
}
\end{figure}

Just like other motors, condensin hydrolyzes ATP, generating $\mu \approx 20\ k_BT$ chemical energy that is converted into mechanical work, which in this case results in extrusion of a DNA loop~\cite{ganji2018real}. To derive an expression for LE velocity, we calculated the thermodynamic work required for LE. The required work $W$ modulates the rate of the mechanical process by the exponential factor $\exp(-W/k_BT)$. In our model, $W$ has two contributions. The first  is the work needed to extrude the DNA at $f=0$ ($W_{0}$). Condensin extrudes the loop by decreasing the spatial distance between the attachment points from $\vR=R_1$ to $\vR=R_2$ (Fig.\ref{fig:modelfig}). 
The associated genomic length of DNA that has to be deformed is $L_{\Sigma} = L_0 + L_1$.  
The second contribution is $W_{f}$, which comes by applying  an external load.  Condensin resists $f$ up to a threshold value~\cite{ganji2018real}, which may be thought of as the stall force. The mechanical work done when condensin takes a step, $\DR=R_1-R_2$, is $W_{f}= f \DR$.

We calculated $W$ as the free energy change needed to bring a semi-flexible polymer with contour length $L_\Sigma$, from the end-to-end distance $R_1$ to $R_2$. It can be estimated using the relation, $W \approx -k_BT\log \big(P(R_2,f|L_\Sigma)\big)+ k_BT\log\big(P(R_1,f|L_\Sigma)\big) $, where $P(\vR,f|\vL)$ is given by $P(\vR,f|\vL) = B P(\vR|\vL)e^{f R/k_BT}$ where $B$ is a  normalization constant.   
Although $P(\vR,f|\vL)$ is a distribution, implying that there is a distribution for $W$,  for illustrative purposes, we plot $W$ in Fig.\ref{fig:delF} for a fixed $R_1=40\nm$ and $R_2=14\nm$ corresponding to $\Delta R= 26\nm$ as estimated using our theory to experiment in later section. It is evident that condensin has to overcome the highest bending penalty in the first step of extrusion, and subsequently $W$ is essentially a constant at large $L_{\Sigma}$. Note that when $f=0$ (red line in Fig.\ref{fig:delF}), $W=W_0$ because the $W_f$ term vanishes.  
If $R_1=40 \nm$, which is approximately the size of the condensin motor, we estimate that condensin pays $5\ k_BT$  to initiate the extrusion process without tension (red line in Fig.\ref{fig:delF}). 

\begin{figure}[]
\centering
\includegraphics[width=\textwidth]{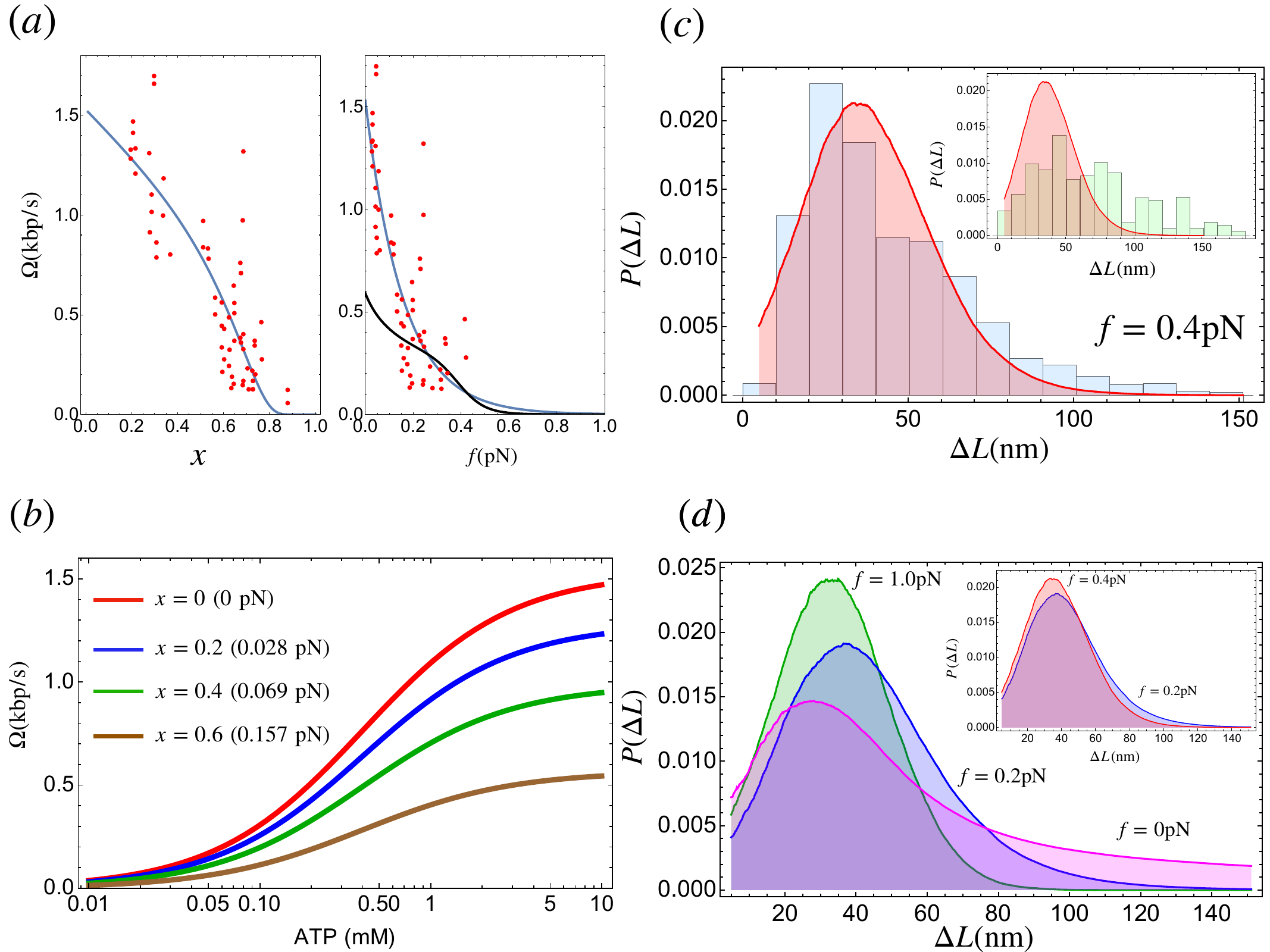}
\caption{\label{fig:fit_model}  (a) {\bf Left panel:} The LE velocity  as a function of the relative extension of the DNA ($x$). Red dots are experimental data~\cite{ganji2018real}, and the solid blue line is the fit obtained using Eq. (\ref{eq:omega}). We used $l_p^{DNA}= 50\nm$ for the persistence length of DNA. {\bf Right panel:} Extrusion velocity as a function of the external load acting on DNA ($f$). The blue line is $\Omega$ from Eq. (\ref{eq:omega}) and the line in black is reproduced from Fig. 6C in~\cite{marko2019dna}. The unit of LE velocity, nm/s, is converted to kbp/s using the conversion $1\bp=0.34\nm$~\cite{alberts2013essential}. (b)  The dependence of LE velocity on ATP concentration  for different relative extension of DNA. The parameters used are $\hat{k}_0=20\persec$, $\DR=26 \nm$, and $K_M=0.4\ \mMol$, where $K_M=0.4\ \mMol$ is Michaelis-menten constant for ATP hydrolysis rate~\cite{terakawa2017condensin}. We used $l_p^{DNA}= 50\nm$ for the persistence length of DNA.
(c)
Distribution of the DNA extrusion length per step ($\Delta L$) using $l^{DNA}_p=42\nm$, which is the value reported in the experiment~\cite{ryu2020resolving}. The histograms are the experimental data taken from Ryu et al.~\cite{ryu2020resolving}(blue) and Strick et al.~\cite{strick2004real}(inset;green). The distributions in red are the theoretical calculations. The distributions for theory and the experiments are both in $f=0.4\pN$.
(d) LE length distributions for various external load on DNA using $l_p^{DNA}= 42\nm$~\cite{ryu2020resolving}. $f=0\pN$ is in magenta, $f=0.2\pN$ is in blue, and $f=1.0\pN$ is in green. The inset compares the results for $f=0.2\pN$ and $f=0.4\pN$.
}
\end{figure}

Once the energetic costs for LE are known, we can calculate the LE velocity as a function of an external load applied to condensin. From energy conservation, we obtain the equality, $n \mu=W_{0} + W_{f} + Q$, where $n$ is the number of ATP molecules consumed per mechanical step, $\mu$ is the energy released by ATP hydrolysis, and $Q$ is the heat dissipated during the extrusion process.  The maximum  force is obtained at equilibrium when the equality $n \mu = W_{0} +  W_{f}\{ f_  {max} \}$ holds.  If we denote the rate of mechanical transition as $k^+$ and reverse rate as $k^-$, fluctuation theorem~\cite{seifert2005fluctuation,seifert2012stochastic,mugnai2020theoretical} together with conservation of energy gives the following relation:  

\begin{align} 
\begin{split}
\label{eq:thermo}
k^{+}/k^{-} = e^{(n \mu-W_{0} - W_{f})/k_BT}. 
\end{split} 
\end{align}
Note that condensin operates in non-equilibrium by transitioning from state 1 to 2, which requires input of energy. The load dependent form for the above equation may be written as,
\begin{align} 
\begin{split}
\label{}
k^{+}= k_0 e^{-W_{f}/k_BT},
\end{split} 
\end{align}
where $k_0=k^{-} e^{(n \mu-W_{0})/k_BT}$ is the rate of the mechanical transition at $0$ load. 
Thus, with  $\DR$ being the extruded length per reaction cycle , the velocity  of LE, $\Omega$, may be written as, 
\begin{align} 
\begin{split}
\label{eq:omega}
\Omega\{f\}= k_0 e^{-f \DR / k_BT}\DR.
\end{split} 
\end{align}
It is worth stating that $k_0$ is the chemical energy dependent term, which includes $\mu$, depends on the nucleotide concentration. In order to obtain the ATP dependence, we assume the Michaelis-Menten form for $k_0$, $\frac{\hat{k}_0[\text{T}]}{[\text{T}]+K_M}$, where [\text{T}] is the concentration of ATP, $\hat{k}_0$ is the maximum rate at saturating ATP concentration, and $K_M$ is the Michaelis-Menten constant (see Fig.\ref{fig:fit_model}b). In principle, $K_M$ should be determined from the measurement of LE velocity as a function of  [\text{T}]. Here, we use $K_M=0.4\ \mMol$ obtained as Michaelis-Menten constant for ATP hydrolysis rate~\cite{terakawa2017condensin}, assuming that ATP hydrolysis rate is the rate-limiting step in the loop extrusion process.


In order to calculate $\Omega$ as a function of the relative DNA extension, $x$, we use the expression ~\cite{marko1995stretching,rubinstein2003polymer},
\begin{align} 
\begin{split}
\label{eq:f}
f = \frac{k_BT}{2l_p}\Big[2x+\frac{1}{2}\Big(\frac{1}{1-x}\Big)^2-\frac{1}{2}\Big].
\end{split} 
\end{align}
The dimensionless variable, $x$, is the $f$-dependent relative extension. In the Ganji et al.~\cite{ganji2018real}  experiment $x$ is measured, and the $f$-dependence of $\Omega$ is obtained by expressing  $x$ in terms of $f$ using a numerical procedure. 


\subsection*{Analysis of experimental data}

\subsubsection*{Loop extrusion velocity}
We used Eq.(\ref{eq:omega}) to fit the experimentally measured LE velocity as a function of DNA extension~\cite{ganji2018real}. The two fitting parameters are $\DR$, and $k_0$, the average step size for condensin, and the extrusion rate at $f=0$, respectively. Excellent  fit of theory to experiments, especially considering the  dispersion in the data, gives  $k_0=20 \persec$ and $\DR=26 \nm$.  This indicates that condensin  undergoes a conformational change that brings the head and the hinge to within $\Delta R \sim 26 \nm$ ($\sim$ 76 bps), during each extrusion cycle. This prediction is remarkably close to the value measured in the recent AFM experiment $\sim 22\nm$~\cite{ryu2020condensin}, and is further supported by our simulations (see below).
We note that $k_0=20 \persec$ is roughly ten times greater than the bulk hydrolysis rate estimated from ensemble experiments~\cite{ganji2018real,terakawa2017condensin}. 
A plausible reason for the apparent discrepancy, already provided in the experimental studies~\cite{ganji2018real,terakawa2017condensin}, is that  bulk hydrolysis rate could underestimate the true hydrolysis rate due to the presence of inactive condensins. Thus, the estimated rate of $k_0 = 2 \persec$ should be viewed as a lower bound~\cite{ganji2018real}. Another possible reason for the discrepancy may be due to methods used to estimate $k_0$ in previous experiments~\cite{ganji2018real,terakawa2017condensin}. It is clear that additional experiments are needed to obtain better estimates of the hydrolysis rate, which is almost all theories is seldom estimated.

In Fig.\ref{fig:fit_model}a right panel we compare the dependence of $\Omega$ on $f$ obtained in a previous kinetic model that has in excess of 20 parameters~\cite{marko2019dna}. In contrast to our theory, even the shape of the LE velocity for $\Omega$ versus $f$ does not agree with experiment. In addition, there is a major discrepancy (factor of 2-3) between the predicted and the measured values of $\Omega$ at low force.

\subsubsection*{LE length distribution}
Recently Ryu et al.~\cite{ryu2020resolving} measured the distribution of LE length per step using magnetic tweezers. We calculated the LE distribution using Eq.(\ref{eq:P(L|R)}) and Eq.(\ref{eq:P(L|R,f)app}).  Of interest here  is the distribution for $L_1-L_2$ where $L_1$ is the captured length of DNA in open shape (see Fig.\ref{fig:modelfig}; right panel). We use the length of condensin in the open state as $R_1=40\nm$ (roughly the peak in the distance between the head and the hinge in the O shape in the wild type condensin), and assume that  $\Delta R=26\nm$ during a single catalytic cycle, as theoretically calculated in the previous section. This gives the length of the closed state, $R_2=14\nm$. Previous experiments~\cite{ryu2020condensin} reported significant fluctuations in the size of the open and closed states, leading to the standard deviation for the change between these two states to be $\Delta=13\nm$. For simplicity, we include the standard deviation for the conformational change in the open state $R_1\pm \Delta$ and keep $R_2$ fixed. This is justified because as $R$ decreases ($R_2 < R_1$), not only does the peak in 
 $P(\vL|\vR)$ and $P_A(\vL|\vR,f)$ moves to small $\vL$ but also there is a decrease in the fluctuation (width of the distribution is smaller) (see Fig.\ref{fig:PLFdist}). This suggests that the variance of the extrusion length owing to $R_2$ is negligible. We assume that the distribution of $R_1$ is a Gaussian centered at $40\nm$ with a standard deviation of $13\nm$. It is reasonable to approximate $L_2 \approx R_2$ since $R_2=14\nm < l^{DNA}_p=50\nm$. With these  parameters in hand, the captured DNA length, Eq.(\ref{eq:P(L|R)}) and Eq.(\ref{eq:P(L|R,f)app}), directly lead to an expression for the distributions for LE length per step. The probability for the LE length per step ($\Delta \vL$) is, $P(\Delta \vL) =  P(\vL+R_2|R_1)$ for $f=0$ and $ P(\Delta \vL | f) = P_{A}(\vL+R_2|R_1,f)$ for $f>0$. 

We compare in Fig.\ref{fig:fit_model} the theoretically calculated distribution for $f=0\pN$ from Eq.(\ref{eq:P(L|R)}) and $f=0.4\pN$ from Eq.(\ref{eq:P(L|R,f)app}) with experiments. The distribution for $f=0 \pN$ cannot be measured using magnetic tweezers but provides insights into the range of LE length that condensin could take at $f=0$.   Remarkably, the calculated distribution is in excellent agreement with the experimental data~\cite{ryu2020resolving}.

\subsection*{Plausible conformational change of condensin in LE process}
\begin{figure}[]
\centering
\includegraphics[width=\textwidth]{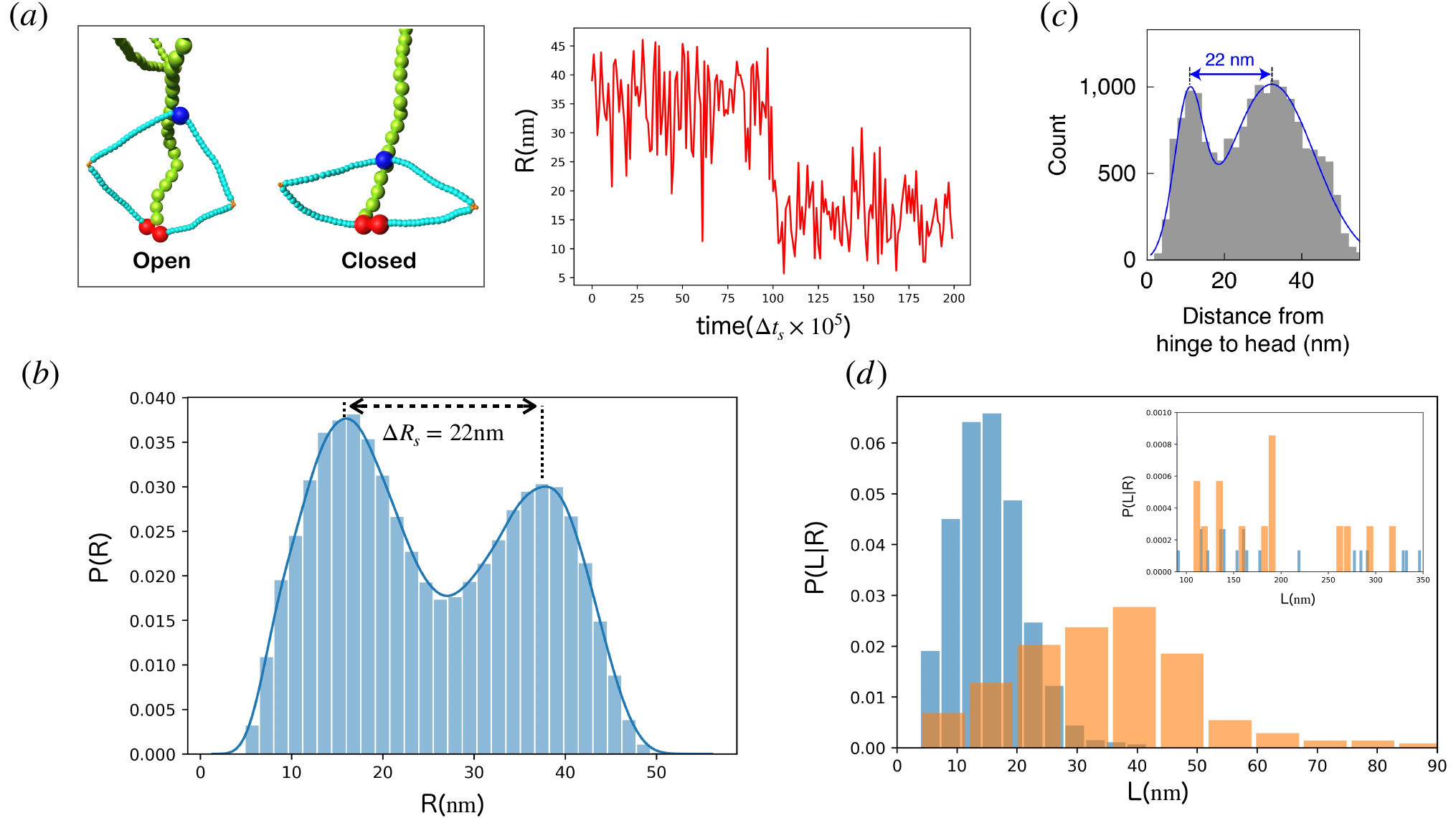}
\caption{\label{fig:delR} Simulations for the transition between O$\rightarrow$B transition. (a) {\bf Left panel:} Representative pictures from the simulation. Red, blue, light blue, orange, and green spheres are the heads, hinge, CC, elbows, and DNA, respectively. {\bf Right panel:} The trajectory for the change in head-hinge distance ($\vR$). $\Delta t_L$ is the time step of simulation (SI Sec.I). (b) Predicted distributions, $P(\vR)$s, of the head-hinge distance during one catalytic cycle of the motor. The peak positions for open state and closed state ($R_1=38\nm$ and $R_2=16\nm$, respectively). The large width and the overlap between the two distributions implies a great deal of conformational heterogeneity that contributes also to broad step size distribution. (c) The distribution for $\vR$ taken from Ryu et al.~\cite{ryu2020condensin}.  The agreement between the simulations and experiments is remarkable, especially considering that the model has {\it no fitting parameters}.
(d) Histograms in orange are the distribution of $\vL$ in the open shape, and the blue histograms are the distribution in the closed shape. The inset shows the distributions for $L>90\nm$. We performed 50 simulations from which 100,000 sample points were used to create the histograms. 
}
\end{figure}

{\bf Simulations without DNA:}
Next we tested whether the predicted value of $\DR\sim 26\nm$, which is in  fair agreement with the experiment, is reasonable using simulations of a simple model.
Because the ATPase domains are located at the heads of condensin, it is natural to assume that the head domain undergoes conformational transitions upon ATP binding and/or hydrolysis. Images of the CCs of the yeast condensin (Smc2-Smc4) using liquid atomic force microscopy (AFM) show they could adopt a few distinct shapes~\cite{eeftens2016condensin,ryu2020condensin}.  Based on these experiments, we hypothesize that the conformational changes initiated at the head domain result in changes in the angle at the junction connecting the motor head to the CC that  propagates through the whole condensin via the CC by an allosteric mechanism~\cite{muir2020structure}. The open (O-shaped in~\cite{ryu2020condensin}), with the hinge that is $\approx 40 \nm$ away from the motor domain, and the closed (B-shape~\cite{ryu2020condensin}) in which the hinge domain is in proximity to the motor domain are the two relevant allosteric states  for LE~\cite{ryu2020condensin,eeftens2016condensin}. To capture the reaction cycle (O $\rightarrow$ B $\rightarrow$ O), we model the CCs as kinked semi-flexible polymers (two moderately stiff segments connected by a flexible elbow), generalizing a similar description of stepping of Myosin V on actin~\cite{hinczewski2013design}. By altering the angle between the two heads the allosteric transition between the open (O-shaped) and closed (B-shaped) states could be simulated (SI contains the details). 

We tracked the head-hinge distance ($\vR$) changes during the transition from the open ($\vR=R_1$) to the closed state ($\vR=R_2$) in order to calculate the distribution of  $\Delta R_s =R_1-R_2$. The sample trajectory in Fig.\ref{fig:delR}a, monitoring the conformational transition between the two states, shows that $\DR_s$ changes by $\sim 22\nm$ for the persistence length of condensin ($l^{CC}_p$; see Sec.II in SI for the detail) $l^{CC}_p = 24 \nm$, which roughly coincides with the value extracted by fitting the theory  to the experimental data.  Higher (smaller) values of $\Delta R_s$ may be obtained using larger (smaller) values of $l_p^{CC}$ (Fig.S3 in SI).
The distributions, $P(\vR)$, calculated from multiple trajectories (Fig.\ref{fig:delR}b) are broad, suggestive of high degree of conformational heterogeneity in the structural transition between the open and closed states. The large dispersions found in the simulations is in surprisingly excellent agreement with experiments~\cite{ryu2020condensin}, which report that the distance between the peaks is $22 \pm  13\nm$.  We find that the corresponding value is $22 \pm 9 \nm$. The uncertainty is calculated using standard deviation in the distributions.  Overall the simulations not only provide insight into the physical basis of the theory but also lend support to recent single molecule experiments~\cite{ryu2020condensin}.

{\bf Simulations with DNA:}
The purpose of the simulations discussed in the previous section was to assess whether the allosteric mechanism produces a structural rationale for the value of $\DR\sim 26\nm$ extracted from the theory.  We also created a simple model of condensin with DNA to give additional insights into the DNA-capture mechanism, which is directly related to the extrusion length of DNA per step. We assume that the capture length of DNA by condensin, $\vL$, is governed by  diffusion of the hinge domain, and that the $\vL$ is solely determined by the semi-flexible polymer nature of DNA.
We attached one end of the DNA to the heads of condensin. The other end of DNA diffuses freely during the simulations. We define the DNA "capture" event by the distance between a DNA segment and the condensin hinge, with a cut-off length of $4\nm$: if the distance is less than $4\nm$ we assume that condensin captures the DNA segment. Captured DNA length is the contour length of DNA held between the heads and the hinge. 
We used a coarse-grained bead-spring model for DNA~\cite{hyeon2006kinetics,dey2017toroidal}. Each bead represents 10 base-pairs, which implies that the bead size is $\sigma_{DNA}=3.4 \nm$. The chain has $N=100$ beads or 1,000 base-pairs. The simulation model for DNA is in the SI (Sec.I).
The simulations with DNA explain an interesting aspect of the DNA capture process. In contrast to well-studied molecular motors that take a step that is nearly constant (conventional kinesin and myosin V) and walks on rigid linear track (microtubule and actin filament, respectively), condensin could in principle capture variable length of DNA during each catalytic cycle because it is a flexible polymer unlike microtubule. 
 Fig.~\ref{fig:delR}d shows that there is a finite probability that $\vL$ exceeds the position of the peak by a considerable amount, as predicted in Fig.~\ref{fig:PLFdist}a. This implies that the $L_1$ can be as large as (60-100) nm ($\sim (180 - 290)  \mathrm{bps}$), which allows for condensin to extrude substantial length of DNA in each catalytic cycle (see the distributions $P(\Delta \vL)$ in Fig.\ref{fig:fit_model}). 

Our results show that theory and simulations for the LE velocity [Eq.(\ref{eq:omega})] predicts the extent of the conformational change of condensin during the LE process fairly accurately ($\sim 26\nm$ in theory and $\sim 22\nm$ in the experiment~\cite{ryu2020condensin} and simulations) and gives the distributions of LE length per cycle in good agreement with experiment. 
Although our theory is in good agreement with experiments for load-dependent LE velocity and distribution of step sizes, the calculated persistence length (24 nm) of the SMC coils is much higher than the value ($\sim$ 3.8 nm) estimated from analyses of AFM images in combination with simulations using the worm-like chain model~\cite{eeftens2016condensin}. It is possible that due to interaction between condensin and other proteins (Brn1 for example) could constrain the head movement, and thus stiffen the coiled coil. However, the discrepancy between theoretical predictions and simulations is  too large to be explained by such effects.  Despite performing many simulations using a variety of polymer models, the origin of this discrepancy is unclear. Simulations with flexible coiled-coil do not  reproduce the measurements of quantities such as $\Delta R$, that are directly monitored in single molecule experiments~\cite{ryu2020condensin} (see Fig.S3). We believe that additional experiments and simulations based on higher resolution structures in different nucleotide states are required to close the gap.

\section*{Discussion}
{\bf Connection to experiments:}
Two of the most insightful experimental studies~\cite{ryu2020condensin,ryu2020resolving} have reported the mean LE velocity and step size distribution as a function of $f$.  Because these are direct single molecule measurements that have caught the motor in the act of LE, the results are unambiguous, requiring little or no interpretation.  Minimally any viable theory must account for these measurements as quantitatively as possible,  using only a small number of physically meaningful parameters. To our knowledge, our theory is currently the only one that reproduces the experimental observations accurately with just two parameters. The only other theory~\cite{marko2019dna} that reported LE velocity as a function of force, not only has a large number of parameters but it cannot be used to calculate the step size distribution.

We first showed that the calculated LE velocity could be fit to experimental data in order to extract the hydrolysis rate and the mean step size. To provide a physical interpretation of the theoretically predicted mean step size, we performed polymer based simulations using a simple model of the CCs. The model for the simulations was based on the AFM images~\cite{ryu2020condensin}, which  showed that, during loop extrusion, there is a transition between the O shape (head and the hinge are far apart) to the B shape where they are closer. Remarkably, our simulations capture the distributions of the head-hinge distances in the O and B states well {\it without any fitting parameters}. The  mean distance between the peaks ($\Delta R_s \approx 22\nm$) in the simulations is in excellent agreement with measured value (see Fig. 2c in~\cite{ryu2020condensin}).  It is worth emphasizing that our theoretical fit to experiment yielded  $\Delta R \sim 26 \nm$, which is also in very good agreement with $\sim22\nm$ measured in the high speed AFM imaging experiment~\cite{ryu2020condensin}. Thus, both experiments and simulations support the mechanism that repeated O$\rightarrow$B shape transitions result in extrusion of the DNA loops. 

{\bf Relation to a previous study:} 
Recently, a four state chemical kinetic model~\cite{marko2019dna}, similar to the ones used to  interpret experiments on stepping of myosin and kinesin motors on polar tracks (actin and microtubule)~\cite{mugnai2020theoretical,kolomeisky2007molecular}, was introduced in order to calculate the $f$-dependent LE velocity and the loop size. The agreement between the predicted dependence on LE velocity and loop size as a function of $f$ is not satisfactory (see Fig. \ref{fig:fit_model}(a)).  Apart from the very large number of parameters (about twenty one in the simplified version of the theory~\cite{marko2019dna}), our two parameter theory differs from the previous study in other important ways. (1) The model~\cite{marko2019dna} is apparently based on the rod-like (or I shape) X-ray structure of the prokaryotic CC of the SMC dimer~\cite{diebold2017structure}, which was pieced together by joining several segments of the CC.  However, the theory itself does not incorporate any structural information but is based on a number of rates connecting the four assumed states in the reaction cycle of the motor, and energetics associated with the isolated DNA.  (2) Because the previous purely kinetic model~\cite{marko2019dna} does not explicitly consider the structure of condensin, it implies that an allosteric communication between the hinge and the head - an integral part of our theory and observed in AFM experiments~\cite{eeftens2016condensin}, is not even considered for the LE mechanism~\cite{marko2019dna}.  The lack of conformational changes in response to ATP-binding implies that the substantial decrease in the head-hinge distance by about $\sim 22\nm$ observed in AFM imaging experiments cannot be explained, as was noted previously~\cite{ryu2020condensin}.  (3) In the picture underlying the DNA capture model~\cite{marko2019dna} (referred to as the DNA pumping model elsewhere~\cite{ryu2020condensin}), the distance between the head and the hinge changes very little, if at all. Such a scenario is explicitly ruled out in an experimental study by Ryu et.  al.,~\cite{ryu2020condensin} in part because they seldom observe the I shape in the holocomplex by itself or in association with DNA. For this reason, we believe that the mechanism proposed in the recent simulation study~\cite{nomidis2021dna} is unlikely to be viable. Rather, it is the O$\leftrightarrow$B transition that drives the loop extrusion process, as found in experiments~\cite{ryu2020condensin}, and affirmed here using our simulations.

{\bf Structural basis for LE:}
The paucity of structures for condensin and cohesin in distinct nucleotide bound states makes it difficult to interpret experiments, theory and simulations in molecular terms.  The situation is further exacerbated because even  the biochemical reaction cycle of condensin (or the related motor cohesin) has not been determined. A recent $8.1\angstrom$ structure of the yeast condensin holocomplex in both the {\it apo} non-engaged state (the one  in which the regulatory element, YCS4, bring the heads in proximity), and the  {\it apo}-bridged state in which the heads interact with each other show a sharp turn in the elbow region, resembling an inverted letter J in the representation  in Fig.1 in a recent study~\cite{lee2020cryo}. The functional importance of the inverted J  state is unclear because when condensin is active (extruding loops in an ATP-dependent manner) only the O- and B- shaped structures are apparently observed~\cite{ryu2020condensin}. Furthermore, the structure of the related motor, cohesin, with DNA shows that the heads are bound to DNA with the hinge is in proximity~\cite{shi2020cryo}, which is inconceivable if the CCs adopt only the I shape.  For this reason, we compared our simulations directly with experiments that have measured the distance changes between the head and the hinge during the {\it active}  LE process~\cite{ryu2020condensin}.  

The partially resolved structure of the ATP-bound state shows a large opening of the CC near the heads~\cite{lee2020cryo}, suggestive of an allosterically driven conformational change.  In contrast, the structure of only the prokaryotic SMC coiled-coil at 3.2 $\angstrom$  resolution, which was created by piecing together several fragments in the CC, showed that it adopts the I shape. As noted elsewhere~\cite{ryu2020condensin}, the I-shaped structure is almost never observed in yeast condensin during its function. The paucity of structures prevents any meaningful inclusion of structural details in theory and simulations. It is for this reason, we resorted to comparisons to AFM imaging data and results from single molecule magnetic tweezer experiments, which have caught the yeast condensin as it executes its function, in validating our theory.  After all it is the function that matters. In the SI (Sec. VI), we performed structural alignment and normal mode analysis using the partially available cryo-EM structures~\cite{lee2020cryo} in order to capture the possible conformational transition in the {\it apo} state that is poised to transition to the LE active state. Even using only partially resolved structures, the normal mode analyses show that ATP binding induces a substantial opening  in the CC region that interacts with the head domains. This preliminary analysis does suggest that for loop extrusion to occur there has to be an allosteric mechanism that brings the head and hinge of the motor close to each other spatially.

{\bf Scrunching versus Translocation:} Using a combination of simulations based on a simple model and theory, we have proposed that LE occurs by a scrunching mechanism. The crux of the scrunching mechanism is that the motor heads, once bound to the DNA, are relatively stationary. Extrusion of the loop occurs by the change in the distance between the head and the hinge by about $\sim 22\nm$. This conformational change is likely driven by ATP binding to condensin~\cite{ryu2020resolving}.  As a result, the head reels in the DNA, with the mean length that could be as large as $\sim 100 \nm$, although the most probable value is $\approx (25-40) \nm$ depending on the external load (Fg. \ref{fig:fit_model}(c) and (d)). Recent experiments have suggested that LE occurs by a scrunching mechanism~\cite{ryu2020resolving}, although it was (as stated earlier) proposed in the context of DNA bubble formation~\cite{kapanidis2006initial}, which is the initial stage in bacterial transcription. The near quantitative agreement with experiments for load-dependent LE velocity and step size distribution shows that the theory and the mechanism are self-consistent. 

In contrast, the other mechanism is based on the picture that condensin must translocate along the DNA in order to extrude loops~\cite{banigan2019limits}. Elsewhere~\cite{ganji2018real} it is argued that the translocation observed in the experiment is an artifact due to salt and buffer conditions, thus casting doubt on the motor translocating along DNA.


{\bf Directionality of LE:}
Directionality is a vital characteristic of biological molecular machines with SMC proteins being no exception. In our model (Fig.~\ref{fig:modelfig}), the unidirectional LE arises during two stages. One is the DNA capture process and the other is the actual active loop extrusion  (State1 to State2 in Fig.~\ref{fig:modelfig}). Directional loop extrusion  emerges as a result of binding  of the SMCs and the associated subunits to DNA via anisotropic interactions. Indeed, the structure of cohesin~\cite{shi2020cryo} suggests that the subunit (STAG1) interacts with the hinge of cohesin by interactions that are anisotropic. This implies that the directional LE could be set by the very act of binding of the condensin motor heads to DNA, which poises the hinge to preferentially interact with  DNA downstream of the motor head, we show schematically as point A in Fig.~\ref{fig:modelfig}.  This in turn ensures  that  the probability of capture of DNA resulting in $\Delta L <0$ is minimized. 

In addition, in our theory there is an asymmetry in the expressions for $k^+$ and $k^-$ in Eq.(4) arising from the free energy, $\mu$, due to the ATP hydrolysis. This process leads to a decrease in the head-hinge distance, as seem in the liquid AFM  images of condensin in action.   However, we believe that, with small probability, $\Delta L<0$ could arise because of the slippage of DNA from the extruded DNA by the strong resistive force on the motor~\cite{ryu2020resolving}. This situation is reminiscent of the slippage of kinesin on microtubule under resistive load~\cite{sudhakar2021germanium}. In the absence of any opposing force, the probability that $\Delta L<0$ is likely to be small.

{\bf Brownian ratchet and power-stroke:}
Whether SMC proteins employ Brownian ratchet mechanism~\cite{astumian1997thermodynamics} or power stroke~\cite{sindelar2002two} is an important question that has to be answered to fully elucidate the molecular mechanism of LE.  A recent study proposed Brownian ratchet model for cohesin~\cite{higashi2021brownian}. An analogy to the well-studied motor conventional kinesin-1 (Kin1), which walks towards the plus end of the stiff microtubule,  is useful. It could be argued that Kin1 makes use of both power stroke and biased diffusion.  Previous studies~\cite{zhang2017parsing,zhang2012dissecting} showed that power stoke (neck-linker docking) propels the trailing head of Kin1 only by $\approx$ 5-6 nm forward, which creates  a strong bias to the next binding site. The rest of the step ($\approx$ 6-8 nm) is completed by diffusion. Therefore, it is possible that both power stroke (sets the directionality)  and Brownian ratchet  are not mutually exclusive. Both the mechanisms could play a role in the LE process as well. Further structural, experimental  and computational studies are  required to resolve fully the interplay between these mechanisms in LE.


\subsection*{Conclusions}
We conclude with a few additional remarks. 
(1) We focused only on one-sided loop extrusion (asymmetric process)  scenario for a single condensin, as demonstrated in the {\it in vitro} experiment~\cite{ganji2018real}.  Whether symmetric LE could occur when more than one condensin loads onto DNA producing Z-loop structures~\cite{kim2020dna}, and if the LE mechanism depends on the species~\cite{kong2020human} is yet to be settled. Similar issues likely exist in loop extrusion mediated by cohesins~\cite{kim2019human,davidson2019dna}. We believe that our work, which only relies on the polymer characteristics of DNA and on an allosteric (action at a distance) mechanism for loop extrusion,  provides a framework for theoretical investigation of LE, accounting for different scenarios. (2) The $f$ dependence of  LE velocity allows us to estimate the time scale for compacting the whole genome.  In particular, if the loop extrusion velocity at $f=0$ is taken to be $\sim$1 kbp/s, we can calculate the LE  time using the following assumptions. The number of condensin I and condensin II that are likely bound to DNA is $\sim$ 3,000 and $\sim$ 500, respectively~\cite{walther2018quantitative}.  Therefore, the loops in the entire chromosome1 ($\sim$ 250 Mbps) could be extruded in a few minutes with the motors operating independently. The assumption that the motors operate independently is reasonable because the linear density (number of motors per genomic base pair) of the bound motors is low. A similar estimate has been made for loop extrusion time by cohesin in the G1 phase of HeLa cells~\cite{davidson2019dna}. These times are faster than the time needed to complete mitosis ($\sim$ an hour)~\cite{gibcus2018pathway}. 
(3) Finally, if LE occurs by scrunching, as gleaned from simulations, and advocated through experimental studies~\cite{ryu2020condensin}, it would imply that the location of the motor is relatively fixed on the DNA and the loop is extruded by ATP-driven shape transitions in the coiled-coils.

\bibliography{mybib.bib}

@article{terakawa2017condensin,
	Author = {Terakawa, Tsuyoshi and Bisht, Shveta and Eeftens, Jorine M and Dekker, Cees and Haering, Christian H and Greene, Eric C},
	Journal = {Science},
	Number = {6363},
	Pages = {672--676},
	Publisher = {American Association for the Advancement of Science},
	Title = {The condensin complex is a mechanochemical motor that translocates along {DNA}},
	Volume = {358},
	Year = {2017}}

@article{ganji2018real,
	Author = {Ganji, Mahipal and Shaltiel, Indra A and Bisht, Shveta and Kim, Eugene and Kalichava, Ana and Haering, Christian H and Dekker, Cees},
	Journal = {Science},
	Number = {6384},
	Pages = {102--105},
	Publisher = {American Association for the Advancement of Science},
	Title = {Real-time imaging of {DNA} loop extrusion by condensin},
	Volume = {360},
	Year = {2018}}

@inproceedings{lawrimore2017rotostep,
  title={RotoStep: a chromosome dynamics simulator reveals mechanisms of loop extrusion},
  author={Lawrimore, Josh and Friedman, Brandon and Doshi, Ayush and Bloom, Kerry},
  booktitle={Cold Spring Harbor symposia on quantitative biology},
  volume={82},
  pages={101--109},
  year={2017},
  organization={Cold Spring Harbor Laboratory Press}
}

@article{diebold2017structure,
  title={Structure of full-length {SMC} and rearrangements required for chromosome organization},
  author={Diebold-Durand, Marie-Laure and Lee, Hansol and Avila, Laura B Ruiz and Noh, Haemin and Shin, Ho-Chul and Im, Haeri and Bock, Florian P and B{\"u}rmann, Frank and Durand, Alexandre and Basfeld, Alrun and others},
  journal={Molecular cell},
  volume={67},
  number={2},
  pages={334--347},
  year={2017},
  publisher={Elsevier}
}

@article{bhattacharjee1997distribution,
  title={Distribution function of the end-to-end distance of semiflexible polymers},
  author={Bhattacharjee, JK and Thirumalai, D and Bryngelson, JD},
  journal={arXiv preprint cond-mat/9709345},
  year={1997}
}

@article{marko1995stretching,
  title={Stretching {DNA}},
  author={Marko, John F and Siggia, Eric D},
  journal={Macromolecules},
  volume={28},
  number={26},
  pages={8759--8770},
  year={1995},
  publisher={ACS Publications}
}

@article{yatskevich2019organization,
  title={Organization of Chromosomal {DNA} by {SMC} Complexes},
  author={Yatskevich, Stanislau and Rhodes, James and Nasmyth, Kim},
  journal={Annual Review of Genetics},
  volume={53},
  pages={445--482},
  year={2019},
  publisher={Annual Reviews}
}

@article{buermann2019folded,
  title={A folded conformation of {MukBEF} and cohesin},
  author={Buermann, Frank and Lee, Byung-Gil and Than, Thane and Sinn, Ludwig and O’Reilly, Francis J and Yatskevich, Stanislau and Rappsilber, Juri and Hu, Bin and Nasmyth, Kim and L{\"o}we, Jan},
  journal={Nature structural \& molecular biology},
  volume={26},
  number={3},
  pages={227},
  year={2019},
  publisher={Nature Publishing Group}
}

@article{marko2019dna,
  title={{DNA}-segment-capture model for loop extrusion by structural maintenance of chromosome {(SMC)} protein complexes},
  author={Marko, John F and De Los Rios, Paolo and Barducci, Alessandro and Gruber, Stephan},
  journal={Nucleic acids research},
  volume={47},
  number={13},
  pages={6956--6972},
  year={2019},
  publisher={Oxford University Press}
}

@article{Wilhelm96PRL,
author={Wilhelm, J and Frey, W},
journal={Phys. Rev. Lett.},
volume={77},
pages={2581},
Year={1996},
}

@article{eeftens2016condensin,
  title={Condensin {Smc2-Smc4} dimers are flexible and dynamic},
  author={Eeftens, Jorine M and Katan, Allard J and Kschonsak, Marc and Hassler, Markus and de Wilde, Liza and Dief, Essam M and Haering, Christian H and Dekker, Cees},
  journal={Cell reports},
  volume={14},
  number={8},
  pages={1813--1818},
  year={2016},
  publisher={Elsevier}
}

@book{rubinstein2003polymer,
  title={Polymer physics},
  author={Rubinstein, Michael and Colby, Ralph H and others},
  volume={23},
  year={2003},
  publisher={Oxford university press New York}
}

@article{seifert2012stochastic,
  title={Stochastic thermodynamics, fluctuation theorems and molecular machines},
  author={Seifert, Udo},
  journal={Reports on progress in physics},
  volume={75},
  number={12},
  pages={126001},
  year={2012},
  publisher={IOP Publishing}
}

@article{fudenberg2016formation,
  title={Formation of chromosomal domains by loop extrusion},
  author={Fudenberg, Geoffrey and Imakaev, Maxim and Lu, Carolyn and Goloborodko, Anton and Abdennur, Nezar and Mirny, Leonid A},
  journal={Cell reports},
  volume={15},
  number={9},
  pages={2038--2049},
  year={2016},
  publisher={Elsevier}
}

@article{nasmyth2001disseminating,
  title={Disseminating the genome: joining, resolving, and separating sister chromatids during mitosis and meiosis},
  author={Nasmyth, Kim},
  journal={Annual Review of Genetics},
  volume={35},
  number={1},
  pages={673--745},
  year={2001},
  publisher={Annual Reviews 4139 El Camino Way, PO Box 10139, Palo Alto, CA 94303-0139, USA}
}

@article{kim2020dna,
  title={{DNA}-loop Extruding Condensin Complexes Can Traverse One Another},
  author={Kim, Eugene and Kerssemakers, Jacob and Shaltiel, Indra and Haering, Christian and Dekker, Cees},
  journal={Biophysical Journal},
  volume={118},
  number={3},
  pages={380a},
  year={2020},
  publisher={Elsevier}
}

@article{kim2019human,
  title={Human cohesin compacts {DNA} by loop extrusion},
  author={Kim, Yoori and Shi, Zhubing and Zhang, Hongshan and Finkelstein, Ilya J and Yu, Hongtao},
  journal={Science},
  volume={366},
  number={6471},
  pages={1345--1349},
  year={2019},
  publisher={American Association for the Advancement of Science}
}

@article{davidson2019dna,
  title={{DNA} loop extrusion by human cohesin},
  author={Davidson, Iain F and Bauer, Benedikt and Goetz, Daniela and Tang, Wen and Wutz, Gordana and Peters, Jan-Michael},
  journal={Science},
  volume={366},
  number={6471},
  pages={1338--1345},
  year={2019},
  publisher={American Association for the Advancement of Science}
}

@article{banigan2019limits,
  title={Limits of chromosome compaction by loop-extruding motors},
  author={Banigan, Edward J and Mirny, Leonid A},
  journal={Physical Review X},
  volume={9},
  number={3},
  pages={031007},
  year={2019},
  publisher={APS}
}

@article{hagstrom2003condensin,
  title={Condensin and cohesin: more than chromosome compactor and glue},
  author={Hagstrom, Kirsten A and Meyer, Barbara J},
  journal={Nature Reviews Genetics},
  volume={4},
  number={7},
  pages={520--534},
  year={2003},
  publisher={Nature Publishing Group}
}

@book{flemming1882zellsubstanz,
  title={Zellsubstanz, kern und zelltheilung},
  author={Flemming, Walther},
  year={1882},
  publisher={Vogel}
}

@book{alberts2013essential,
  title={Essential cell biology},
  author={Alberts, Bruce and Bray, Dennis and Hopkin, Karen and Johnson, Alexander D and Lewis, Julian and Raff, Martin and Roberts, Keith and Walter, Peter},
  year={2013},
  publisher={Garland Science}
}

@article{seifert2005fluctuation,
  title={Fluctuation theorem for a single enzym or molecular motor},
  author={Seifert, U},
  journal={EPL {(Europhysics Letters)}},
  volume={70},
  number={1},
  pages={36},
  year={2005},
  publisher={IOP Publishing}
}

@article{eastman2017openmm,
  title={OpenMM 7: Rapid development of high performance algorithms for molecular dynamics},
  author={Eastman, Peter and Swails, Jason and Chodera, John D and McGibbon, Robert T and Zhao, Yutong and Beauchamp, Kyle A and Wang, Lee-Ping and Simmonett, Andrew C and Harrigan, Matthew P and Stern, Chaya D and others},
  journal={PLoS computational biology},
  volume={13},
  number={7},
  pages={e1005659},
  year={2017},
  publisher={Public Library of Science}
}

@article{hyeon2006kinetics,
  title={Kinetics of interior loop formation in semiflexible chains},
  author={Hyeon, Changbong and Thirumalai, D},
  journal={The Journal of Chemical Physics},
  volume={124},
  number={10},
  pages={104905},
  year={2006},
  publisher={American Institute of Physics}
}

@article{strick2004real,
  title={Real-time detection of single-molecule {DNA} compaction by condensin {I}},
  author={Strick, Terence R and Kawaguchi, Tatsuhiko and Hirano, Tatsuya},
  journal={Current biology},
  volume={14},
  number={10},
  pages={874--880},
  year={2004},
  publisher={Elsevier}
}

@article{kremer1991erratum,
  title={Erratum: Dynamics of entangled polymer melts: A molecular-dynamics simulation [J. Chem. Phys. 9 2, 5057 (1990)]},
  author={Kremer, Kurt and Grest, Gary S},
  journal={The Journal of Chemical Physics},
  volume={94},
  number={5},
  pages={4103--4103},
  year={1991},
  publisher={American Institute of Physics}
}

@article{alipour2012self,
  title={Self-organization of domain structures by {DNA}-loop-extruding enzymes},
  author={Alipour, Elnaz and Marko, John F},
  journal={Nucleic acids research},
  volume={40},
  number={22},
  pages={11202--11212},
  year={2012},
  publisher={Oxford University Press}
}

@article{goloborodko2016compaction,
  title={Compaction and segregation of sister chromatids via active loop extrusion},
  author={Goloborodko, Anton and Imakaev, Maxim V and Marko, John F and Mirny, Leonid},
  journal={Elife},
  volume={5},
  pages={e14864},
  year={2016},
  publisher={eLife Sciences Publications Limited}
}

@article{kapanidis2006initial,
  title={Initial transcription by {RNA} polymerase proceeds through a {DNA}-scrunching mechanism},
  author={Kapanidis, Achillefs N and Margeat, Emmanuel and Ho, Sam On and Kortkhonjia, Ekaterine and Weiss, Shimon and Ebright, Richard H},
  journal={Science},
  volume={314},
  number={5802},
  pages={1144--1147},
  year={2006},
  publisher={American Association for the Advancement of Science}
}

@article{chen2010promoter,
  title={Promoter melting triggered by bacterial {RNA} polymerase occurs in three steps},
  author={Chen, Jie and Darst, Seth A and Thirumalai, D},
  journal={Proceedings of the National Academy of Sciences},
  volume={107},
  number={28},
  pages={12523--12528},
  year={2010},
  publisher={National Acad Sciences}
}

@article{thirumalai1999time,
  title={Time scales for the formation of the most probable tertiary contacts in proteins with applications to cytochrome c},
  author={Thirumalai, D},
  journal={The Journal of Physical Chemistry B},
  volume={103},
  number={4},
  pages={608--610},
  year={1999},
  publisher={ACS Publications}
}

@article{mugnai2020theoretical,
  title={Theoretical perspectives on biological machines},
  author={Mugnai, Mauro L and Hyeon, Changbong and Hinczewski, Michael and Thirumalai, D},
  journal={Reviews of Modern Physics},
  volume={92},
  number={2},
  pages={025001},
  year={2020},
  publisher={APS}
}

@article{hinczewski2013design,
  title={Design principles governing the motility of myosin {V}},
  author={Hinczewski, Michael and Tehver, Riina and Thirumalai, D},
  journal={Proceedings of the National Academy of Sciences},
  volume={110},
  number={43},
  pages={E4059--E4068},
  year={2013},
  publisher={National Acad Sciences}
}

@article{zhang2017parsing,
  title={Parsing the roles of neck-linker docking and tethered head diffusion in the stepping dynamics of kinesin},
  author={Zhang, Zhechun and Goldtzvik, Yonathan and Thirumalai, Dave},
  journal={Proceedings of the National Academy of Sciences},
  volume={114},
  number={46},
  pages={E9838--E9845},
  year={2017},
  publisher={National Acad Sciences}
}

@article{dey2017toroidal,
  title={Toroidal condensates by semiflexible polymer chains: Insights into nucleation, growth and packing defects},
  author={Dey, Atreya and Reddy, Govardhan},
  journal={The Journal of Physical Chemistry B},
  volume={121},
  number={39},
  pages={9291--9301},
  year={2017},
  publisher={ACS Publications}
}

@article{ryu2020condensin,
  title={The condensin holocomplex cycles dynamically between open and collapsed states},
  author={Ryu, Je-Kyung and Katan, Allard J and van der Sluis, Eli O and Wisse, Thomas and de Groot, Ralph and Haering, Christian H and Dekker, Cees},
  journal={Nature structural \& molecular biology},
  pages={1--8},
  year={2020},
  publisher={Nature Publishing Group}
}

@article{lee2020cryo,
  title={Cryo-EM structures of holo condensin reveal a subunit flip-flop mechanism},
  author={Lee, Byung-Gil and Merkel, Fabian and Allegretti, Matteo and Hassler, Markus and Cawood, Christopher and Lecomte, L{\'e}a and O’Reilly, Francis J and Sinn, Ludwig R and Gutierrez-Escribano, Pilar and Kschonsak, Marc and others},
  journal={Nature Structural \& Molecular Biology},
  volume={27},
  number={8},
  pages={743--751},
  year={2020},
  publisher={Nature Publishing Group}
}

@article{muir2020structure,
  title={The structure of the cohesin {ATP}ase elucidates the mechanism of {SMC}--{kleisin} ring opening},
  author={Muir, Kyle W and Li, Yan and Weis, Felix and Panne, Daniel},
  journal={Nature Structural \& Molecular Biology},
  volume={27},
  number={3},
  pages={233--239},
  year={2020},
  publisher={Nature Publishing Group}
}

@article{shi2020cryo,
  title={Cryo-EM structure of the human cohesin-{NIPBL}-{DNA} complex},
  author={Shi, Zhubing and Gao, Haishan and Bai, Xiao-chen and Yu, Hongtao},
  journal={Science},
  year={2020},
  publisher={American Association for the Advancement of Science}
}

@article{roberts2006multiseq,
  title={MultiSeq: unifying sequence and structure data for evolutionary analysis},
  author={Roberts, Elijah and Eargle, John and Wright, Dan and Luthey-Schulten, Zaida},
  journal={BMC bioinformatics},
  volume={7},
  number={1},
  pages={382},
  year={2006},
  publisher={Springer}
}

@article{ryu2020resolving,
  title={Resolving the step size in condensin-driven {DNA} loop extrusion identifies {ATP} binding as the step-generating process},
  author={Ryu, Je-Kyung and Rah, Sang-Hyun and Janissen, Richard and Kerssemakers, Jacob WJ and Dekker, Cees},
  journal={Available at SSRN 3728949},
  year={2020}
}

@article{kolomeisky2007molecular,
  title={Molecular motors: a theorist's perspective},
  author={Kolomeisky, Anatoly B and Fisher, Michael E},
  journal={Annu. Rev. Phys. Chem.},
  volume={58},
  pages={675--695},
  year={2007},
  publisher={Annual Reviews}
}

@article{mugnai2020role,
  title={Role of Long-range Allosteric Communication in Determining the Stability and Disassembly of SARS-COV-2 in Complex with ACE2},
  author={Mugnai, Mauro Lorenzo and Templeton, Clark and Elber, Ron and Thirumalai, Dave},
  journal={bioRxiv},
  year={2020},
  publisher={Cold Spring Harbor Laboratory}
}

@article{gibcus2018pathway,
  title={A pathway for mitotic chromosome formation},
  author={Gibcus, Johan H and Samejima, Kumiko and Goloborodko, Anton and Samejima, Itaru and Naumova, Natalia and Nuebler, Johannes and Kanemaki, Masato T and Xie, Linfeng and Paulson, James R and Earnshaw, William C and others},
  journal={Science},
  volume={359},
  pages={eaao6135},
  number={6376},
  year={2018},
  publisher={American Association for the Advancement of Science}
}

@article{walther2018quantitative,
  title={A quantitative map of human Condensins provides new insights into mitotic chromosome architecture},
  author={Walther, Nike and Hossain, M Julius and Politi, Antonio Z and Koch, Birgit and Kueblbeck, Moritz and {\O}deg{\aa}rd-Fougner, {\O}yvind and Lampe, Marko and Ellenberg, Jan},
  journal={Journal of Cell Biology},
  volume={217},
  number={7},
  pages={2309--2328},
  year={2018},
  publisher={Rockefeller University Press}
}

@article{kong2020human,
  title={Human condensin {I} and {II} drive extensive {ATP}-dependent compaction of nucleosome-bound {DNA}},
  author={Kong, Muwen and Cutts, Erin E and Pan, Dongqing and Beuron, Fabienne and Kaliyappan, Thangavelu and Xue, Chaoyou and Morris, Edward P and Musacchio, Andrea and Vannini, Alessandro and Greene, Eric C},
  journal={Molecular cell},
  volume={79},
  number={1},
  pages={99--114},
  year={2020},
  publisher={Elsevier}
}

@article{chiu2004dna,
  title={DNA interaction and dimerization of eukaryotic SMC hinge domains},
  author={Chiu, Allen and Revenkova, Ekaterina and Jessberger, Rolf},
  journal={Journal of Biological Chemistry},
  volume={279},
  number={25},
  pages={26233--26242},
  year={2004},
  publisher={Elsevier}
}

@article{griese2010structure,
  title={Structure and DNA binding activity of the mouse condensin hinge domain highlight common and diverse features of SMC proteins},
  author={Griese, Julia J and Witte, Gregor and Hopfner, Karl-Peter},
  journal={Nucleic acids research},
  volume={38},
  number={10},
  pages={3454--3465},
  year={2010},
  publisher={Oxford University Press}
}

@article{alt2017specialized,
  title={Specialized interfaces of Smc5/6 control hinge stability and DNA association},
  author={Alt, Aaron and Dang, Hung Q and Wells, Owen S and Polo, Luis M and Smith, Matt A and McGregor, Grant A and Welte, Thomas and Lehmann, Alan R and Pearl, Laurence H and Murray, Johanne M and others},
  journal={Nature communications},
  volume={8},
  number={1},
  pages={1--14},
  year={2017},
  publisher={Nature Publishing Group}
}

@article{astumian1997thermodynamics,
  title={Thermodynamics and kinetics of a Brownian motor},
  author={Astumian, R Dean},
  journal={science},
  volume={276},
  number={5314},
  pages={917--922},
  year={1997},
  publisher={American Association for the Advancement of Science}
}

@article{sindelar2002two,
  title={Two conformations in the human kinesin power stroke defined by X-ray crystallography and EPR spectroscopy},
  author={Sindelar, Charles V and Budny, Mary Jane and Rice, Sarah and Naber, Nariman and Fletterick, Robert and Cooke, Roger},
  journal={Nature structural biology},
  volume={9},
  number={11},
  pages={844--848},
  year={2002},
  publisher={Nature Publishing Group}
}

@article{higashi2021brownian,
  title={A Brownian ratchet model for DNA loop extrusion by the cohesin complex},
  author={Higashi, Torahiko L and Tang, Minzhe and Pobegalov, Georgii and Uhlmann, Frank and Molodtsov, Maxim},
  journal={bioRxiv},
  year={2021},
  publisher={Cold Spring Harbor Laboratory}
}

@article{zhang2012dissecting,
  title={Dissecting the kinematics of the kinesin step},
  author={Zhang, Zhechun and Thirumalai, D},
  journal={Structure},
  volume={20},
  number={4},
  pages={628--640},
  year={2012},
  publisher={Elsevier}
}

@article{sudhakar2021germanium,
  title={Germanium nanospheres for ultraresolution picotensiometry of kinesin motors},
  author={Sudhakar, Swathi and Abdosamadi, Mohammad Kazem and Jachowski, Tobias J{\"o}rg and Bugiel, Michael and Jannasch, Anita and Sch{\"a}ffer, Erik},
  journal={Science},
  volume={371},
  number={6530},
  year={2021},
  publisher={American Association for the Advancement of Science}
}

@article{sanborn2015chromatin,
  title={Chromatin extrusion explains key features of loop and domain formation in wild-type and engineered genomes},
  author={Sanborn, Adrian L and Rao, Suhas SP and Huang, Su-Chen and Durand, Neva C and Huntley, Miriam H and Jewett, Andrew I and Bochkov, Ivan D and Chinnappan, Dharmaraj and Cutkosky, Ashok and Li, Jian and others},
  journal={Proceedings of the National Academy of Sciences},
  volume={112},
  number={47},
  pages={E6456--E6465},
  year={2015},
  publisher={National Acad Sciences}
}

@article{wilhelm1996radial,
  title={Radial distribution function of semiflexible polymers},
  author={Wilhelm, Jan and Frey, Erwin},
  journal={Physical review letters},
  volume={77},
  number={12},
  pages={2581},
  year={1996},
  publisher={APS}
}

@article{nomidis2021dna,
  title={{DNA} tension-modulated translocation and loop extrusion by {SMC} complexes revealed by molecular dynamics simulations},
  author={Nomidis, Stefanos K and Carlon, Enrico and Gruber, Stephan and Marko, John F},
  journal={bioRxiv},
  year={2021},
  publisher={Cold Spring Harbor Laboratory}
}



\section*{Acknowledgements}
We thank Rasika Harshey, Changbong Hyeon, Mauro Mugnai, and Johannes Stigler for useful comments and discussions. We are grateful to John Marko for clarifying some aspects of his model.  This work was supported by NSF (CHE 19-00093), NIH (GM - 107703) and the Welch Foundation Grant F-0019 through the Collie-Welch chair. 

\section*{Author contributions statement}

R.T and D.T. conceived the theory,  R.T.,  A.D., G.S., and D.T performed the calculations and simulations, the experiment(s), R.T., A.D., FG. S., and D. T.  analyzed the results.  All authors reviewed the manuscript. 







\end{document}



\title{Supplementary Information \\Theory and Simulations of condensin mediated loop extrusion in DNA}



\author{Ryota Takaki}
\affiliation{Department of Physics, The university of Texas at Austin}
\author{Atreya Dey}
\affiliation{Department of Chemistry, The university of Texas at Austin}
\author{Guang Shi}
\affiliation{Department of Chemistry, The university of Texas at Austin}
\author{D. Thirumalai}

\affiliation{Department of Chemistry, The university of Texas at Austin}

\date{\today}



\maketitle

\tableofcontents
\newpage

\section{Simulations}
\begin{figure}[]
\centering
\includegraphics[width=1\textwidth]{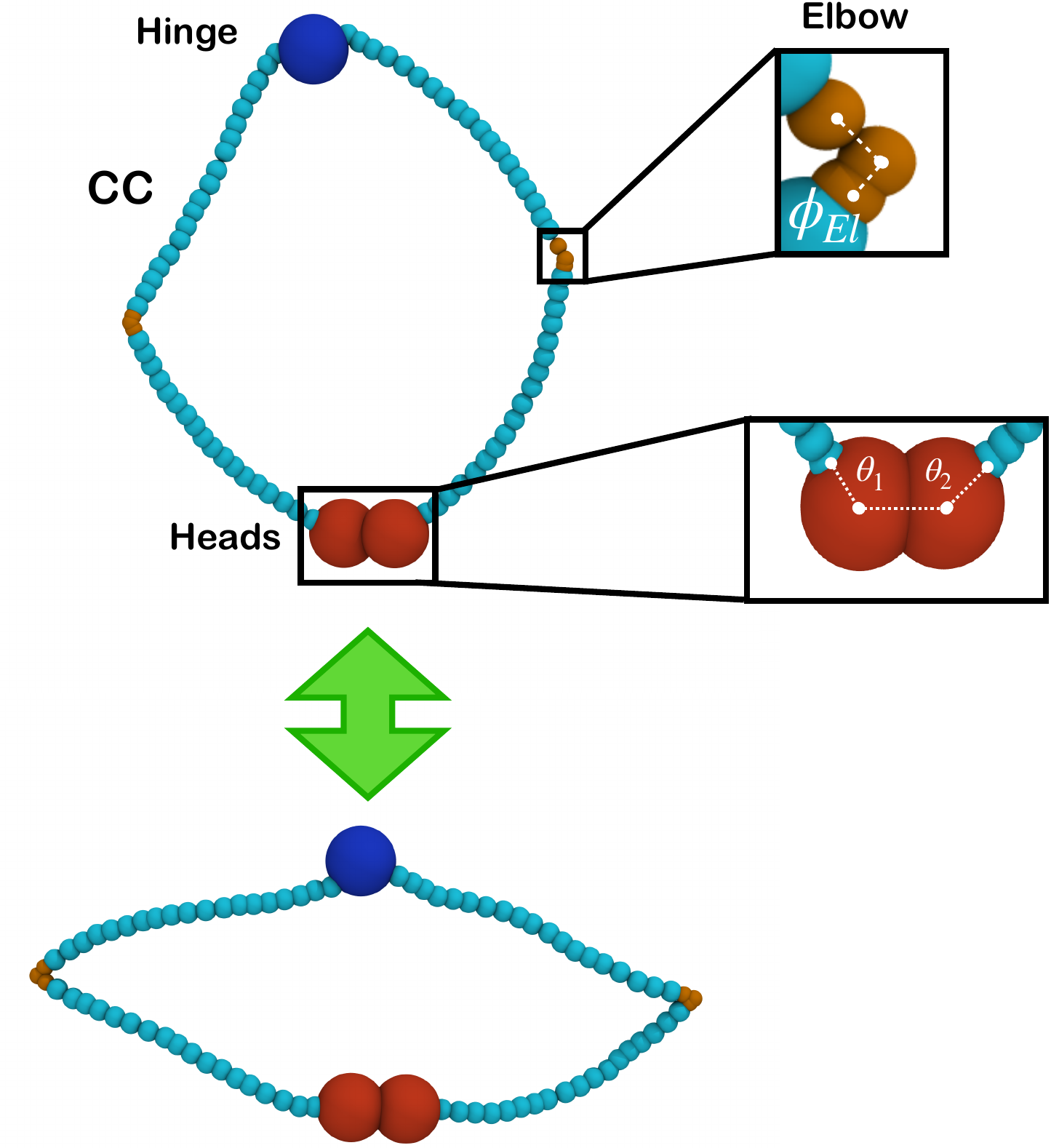}
\caption{\label{fig:simulation} Cartoon representation of the condensin motor based in part on partially resolved structures and FM images.  The simulation model is based on this representation. The two heads are shown as red spheres. A magnified image of the angle between the motor heads at the junctions to the two arms of the SMC is shown in the lower box. The angle at the elbow is depicted in the upper box. The coiled coils (CCs) connecting the motor to the hinge (purple sphere) are treated as a semi-flexible polymers that are kinked  at the flexible elbow region. We envision that the allosteric transition between the open and the closed states (shown as by the green arrow) is driven by ATP binding to the motor domains. 
}
\end{figure}
The purpose of the simulations using a simplified model is to show that the extracted parameter values, obtained by fitting the theoretical extrusion rate or equivalently the velocity of loop extrusion (LE) to experiments, are reasonable. In particular, we use simulations to argue that the value of $\Delta R \approx$ 26 nm or equivalently $\sim$ 76 bps (see the main text for details) is consistent with experiments that have imaged the shape transitions during the LE process. To this end, we imagine that during the ATPase cycle, the SMC motor undergoes a conformational change from an "open" (top structure in Fig. \ref{fig:simulation}) to a "closed"  (bottom structure) state, thus decreasing  the distance between the motor domains and the hinge region. We theorize that this process is allosterically driven, in a manner similar to other cargo-carrying  motors (myosins, kinesins and dynein), by binding ATP. We envision that in the SMC the allosteric transitions are effectuated through the movement of the flexible elbow region. Such a picture is consistent with high speed AFM imaging~\cite{eeftens2016condensin} and measurements of head-hinge distance as the motor transitions between the two active (open and closed) states~\cite{ryu2020condensin}. 

\subsection{Model for Condensin \label{sec:model}}
We modeled the two heads of condensin as spheres that are connected by finitely-extensible nonlinear elastic (FENE) potential~\cite{kremer1991erratum} to the coiled coils (CCs) that connect the motor domains to the hinge (Fig.\ref{fig:simulation}). The CC in the SMCs are reminiscent of the lever arm in Myosin V. The angle $\theta_1$ and $\theta_2$ are formed at the junctions connecting the motor heads to the first bead on the CCs (Fig.\ref{fig:simulation}).  As in the well-studied molecular motors, a  change in the conformation  initiated in the head domain, is  amplified to the rest of the motor through the CCs.  We envision this process as the principle mechanism by which a spool (roughly $\frac{\Delta R}{0.34}=$ 76 base pairs (bps)  in a single step) of DNA could be extruded. 

We used 19 and 18 beads for upper CC and lower CC, respectively. The diameter of each bead is $1 \nm$ diameter. We used 3 beads (diameter $0.4\nm$ each) in the middle of the CCs for the elbow region. The strength of the angle potential in the elbow region, marking the break in an otherwise stiff CC, is chosen to facilitate the allosteric propagation of conformational changes in the motor head. For the hinge and two motor heads we used $4\nm$ diameter beads. 

All the lengths are measured in units of $\sigma=1\nm$, corresponding to the diameter of the beads in the CC. We express energy in  $k_BT$ units, where $k_B$ is the Boltzmann constant and $T$ is the temperature. The mass of all the particles were set to $m=1$. We performed low-friction Langevin dynamics simulations using OpenMM \cite{eastman2017openmm} software using a  time-step of $\Delta t_L=0.01\tau_L$, where $\tau_L=0.4\sqrt{m\sigma^2/k_BT}$. The value of the friction coefficient is $0.01/\tau_L$. 

\subsection{Energy function} Because our goal is to merely illustrate that the proposed allosteric mechanism for SMC-mediated LE is plausible, we chose a simple energy function to monitor the conformational changes in condensin. The explicit form of the energy function is,
\begin{align} 
\begin{split}
\label{eq:Hamil}
E(\Vec{r}_1,\Vec{r}_2,...,\Vec{r}_N, \vec{\phi},\vec{\theta})  &= \sum_{i=1}^{N-1}U_{FENE}(r_{i,i+1}) + \sum_{i\neq j}^{N}U_{N}(r_{i,j}) +\Big( \sum_{i\in CC\neq El}U^{CC }_{ANG}(\phi_{i}) + \sum_{i\in El}U^{El}_{ANG}(\phi_{i}) \Big) \\
& + \sum_{i\in Head}^{}U_{CNF}^{}(\theta_{i}) .
\end{split}  
\end{align}
The first term in Eq.(\ref{eq:Hamil}) enforces the connectivity of the beads and is given by,
\begin{align} 
\begin{split}
\label{}
U_{FENE}(r_{i,i+1}) = -\frac{1}{2}k_F R_F^2 \log\Big[1-\frac{(r_{i,i+1}-r^0_{i,i+1})^2}{R_F^2}\Big],
\end{split}  
\end{align}
where $k_F$ is the stiffness of the potential, $R_F$ is the upper bound for the displacement, and $r^0_{i,i+1}$ is the equilibrium distance between the beads, $i$ and $i+1$.
The second term in Eq.(\ref{eq:Hamil}), accounting for excluded volume interactions, is given by,
\begin{equation} 
\label{}
U_{N}(r_{i,j})  =\epsilon_{N}\Big (\frac{\sigma}{r_{i,j}}\Big )^{12},
\end{equation}
where $\epsilon_{N}$ and $\sigma$  are the strength and range of the interaction, respectively. We used additive interactions, which means that $\sigma$ is  the sum of the radii of the two interacting beads. 
The third and the fourth terms in Eq.(\ref{eq:Hamil}) are the two angle potentials that control the bending stiffness of the CCs. The potential, $U^{}_{ANG}(\phi_{i})$, is taken to be, 
\begin{equation} 
\label{}
U^{}_{ANG}(\phi_{i})  =\epsilon _b(1+\cos \phi_i),
\end{equation}
where $\epsilon_b$,  the energy scale for bending, is related to the persistence length of the semi-flexible CC.
We used a smaller value of $\epsilon_b$ for $U^{El}_{ANG}(\epsilon_b^{El})$ to model the difference in the stiffness between the elbow region, and the rest of the CC. 

The last term in Eq.(\ref{eq:Hamil}) models the conformation change in the motor head  due to ATP binding, and is taken as, 
\begin{equation} 
\label{}
U_{CNF}^{}(\theta_{i})= \frac{1}{2}k_C\big(\theta^0_i-\theta_i\big)^2,
\end{equation}
where $k_C$ is the spring constant for the potential, and $\theta^0_i$ is the equilibrium angle for the angle potential. Before the conformational change, we set $\theta_i^0=2.4$ (radian) in the open state, which is roughly the angle calculated by ATP engaged state of  prokaryotic SMC~\cite{diebold2017structure}. Because the structure for the closed state is unavailable, we chose $\theta_i^0=4.0$ (radian) for the closed state, which leads to $\theta_i\sim \pi$ (radian) in equilibrium. The transition between the open and closed states results in the scrunching of the DNA (the nearly stationary motor reels in the DNA), and extrusion of the loop.
The parameter values in the energy function used in the simulations are in Table~\ref{Table:parameters}. 

\subsection{Simulation with DNA}
We used a coarse-grained bead-spring model for DNA~\cite{hyeon2006kinetics,dey2017toroidal}. Each bead represents 10 base-pairs, which implies that the bead size is $\sigma_{DNA}=3.4 \nm$. The chain has $N=100$ beads or 1,000 base-pairs. Consecutive beads along the chain are  connected by a FENE potential (Eq.(S2)) with $r^0_{i,i+1}=\sigma_{DNA}$. The DNA stiffness is modeled using a harmonic bending potential given by,
\begin{equation}
    U_{BEND} = \frac{1}{2}\sum_{i=1}^{N_{ang}}\alpha\cdot\theta^2,
\end{equation}
where $N_{ang}=98$ is the number of bond angles, $\theta_i$ is the deviation of the $i^{th}$ bond-angle in the chain from $180\deg$, and $\alpha = 15.55$ $(k_B T /\text{rad}^2)$ is a constant~\cite{hyeon2006kinetics}. We chose  $\alpha$ so that the persistence length is $\approx 50\nm$, which is the canonical value for DNA in monovalent salts.   

\begin{table}[t]
\begin{center}
  \begin{tabular}{|c | c|}
    \hline
    \text{Parameter} &\text{Value}\\
    \hline \hline
    $k_F$      &$50 (k_BT/\mathrm{nm}^2)$\\ 
    $R_F$    &$1.5 (\mathrm{nm})$\\
    $\epsilon_b^{CC}$ &$150 (k_BT)$\\
    $\epsilon_b^{El}$ &$2 (k_BT)^*$\\
    $\epsilon_{N}$    &$5(k_BT)$\\
    $k_C$&$200(k_BT/{\text{rad}}^2)$\\
    \hline
  \end{tabular}
\end{center} 
\caption{\label{Table:parameters}Parameters for the molecular dynamics simulation. $^*$ The dependence of the distance distribution between the head and hinge during one cycle on   $\epsilon_b^{El}$ is given in Fig. \ref{fig:dists_lp}.}
\end{table}

\section{Persistence length of condensin}
\begin{figure}[]
\centering
\includegraphics[width=\textwidth]{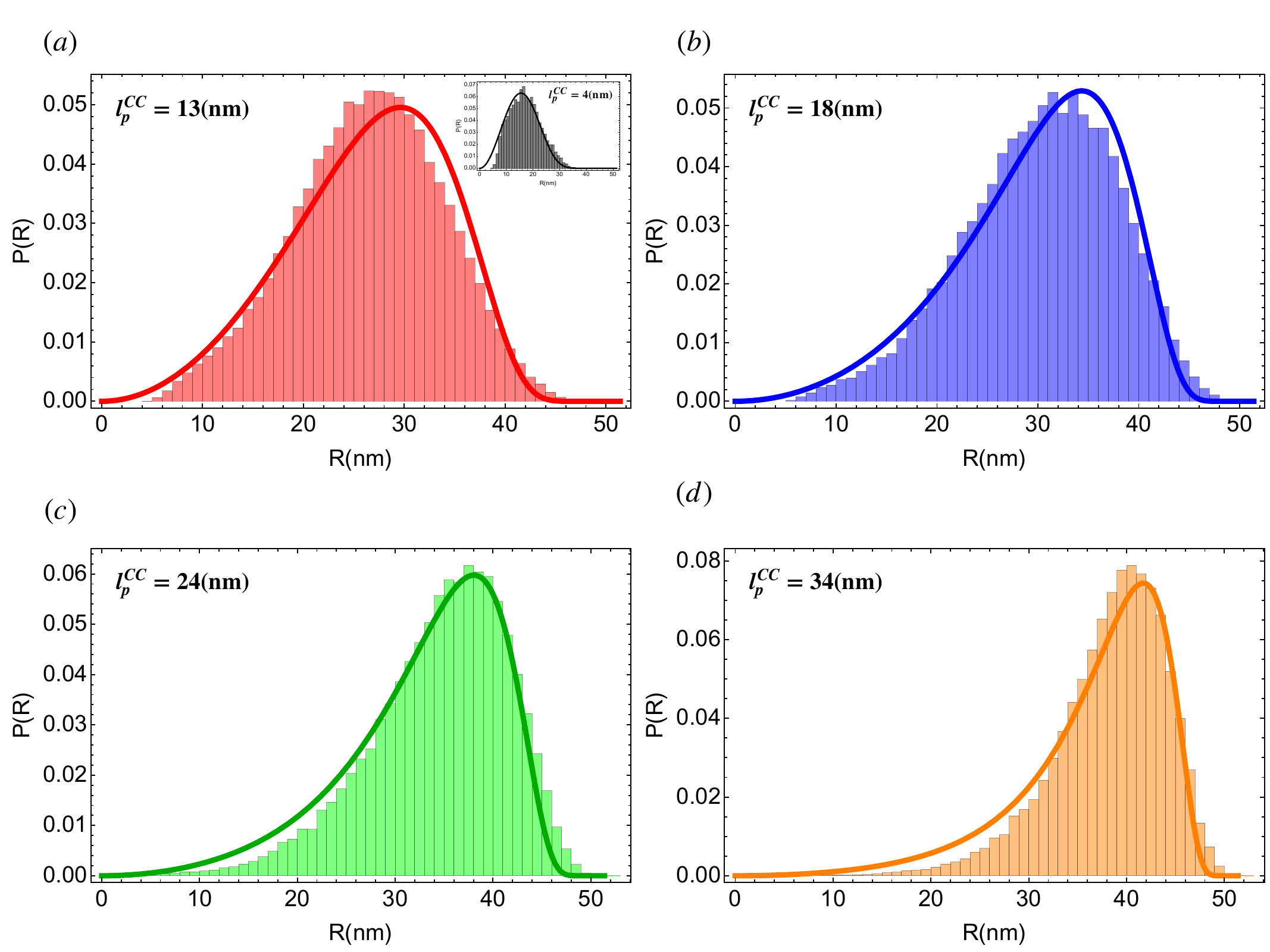}
\caption{\label{fig:persistence} Extracting $l_p^{CC}$ from the simulations. Histograms are for the head-hinge distance in the O shape from the  simulations. Solid lines are from the theoretical expression for $P(R)$. The value of $\epsilon^{CC}_b=150 (k_BT)$ in all the panels except in the inset of (a). (a) $l_p^{CC}\sim13\nm$ for $\epsilon^{El}_b=0 (k_BT)$.  Inset: $l_p^{CC}\sim4\nm$ for $\epsilon^{CC}_b=4 (k_BT)$ and $\epsilon^{El}_b=4 (k_BT)$. (b) $l_p^{CC}\sim18\nm$ and $\epsilon^{El}_b=1 (k_BT)$. (c) $l_p^{CC}\sim24\nm$  and $\epsilon^{El}_b=2 (k_BT)$. (d) $l_p^{CC}\sim34\nm$  and $\epsilon^{El}_b=3 (k_BT)$. 
}
\end{figure}

\begin{figure}[]
\centering
\includegraphics[width=\textwidth]{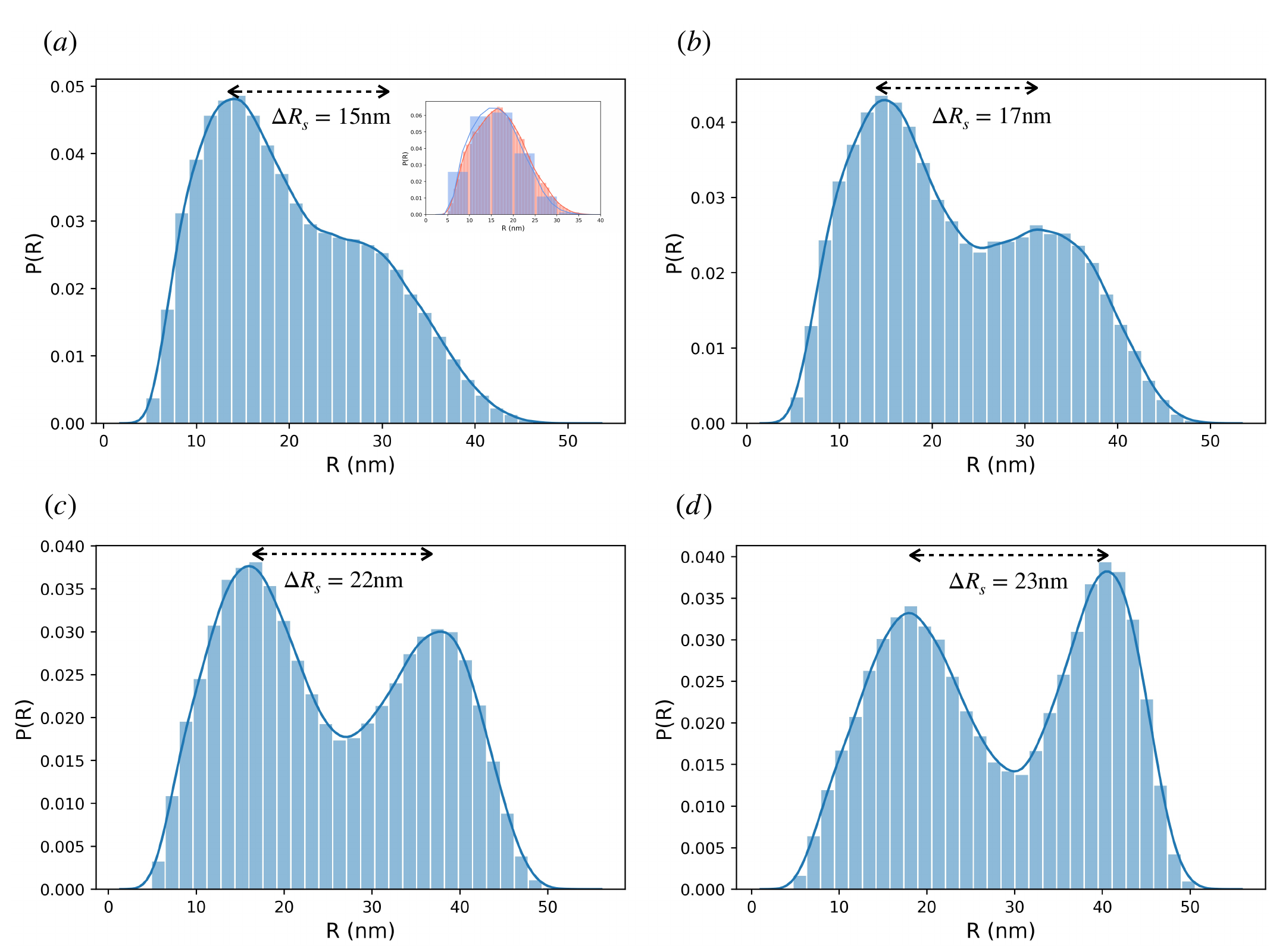}
\caption{\label{fig:dists_lp} $P(R)$ for different values of $l_p^{CC}$ in the simulations. (a) $l_p^{CC}\sim13\nm$ [$\epsilon^{El}_b=0 (k_BT)$]. The position of the peaks are $R_2 \sim 13 \nm$ and $R_1 \sim 28 \nm$.  Inset: $l_p^{CC}\sim4\nm$ [$\epsilon^{CC}_b=\epsilon^{El}_b=4 (k_BT)$].  The position of the peaks are $R_2 \sim 15 \nm$ and $R_1 \sim 16 \nm$. (b)  $l_p^{CC}\sim 18\nm$ [$\epsilon^{El}_b=1 (k_BT)$]. The position of the peaks are $R_2 \sim 14 \nm$ and $R_1 \sim 32 \nm$.  (c) $l_p^{CC}\sim24\nm$ [$\epsilon^{El}_b=2 (k_BT)$]. The position of the peaks are $R_2 \sim 16 \nm$ and $R_1\sim 38 \nm$.  (d)  $l_p^{CC}\sim34\nm$ [$\epsilon^{El}_b=4 (k_BT)$]. The peak positions are $R_2 \sim 18 \nm$ and $R_1\sim 40 \nm$. 
}
\end{figure}

In order for the allosteric transition that brings the head-hinge distance to within $\Delta R\sim 26\nm$ ($\approx 22\nm$ in experiments and simulations), condensin has to be sufficiently rigid but not overly so. 
We estimate the persistence length of condensin ($l^{CC}_p$) by fitting the theoretical expression for end-to-end distance ($R$) with contour length $L$~\cite{bhattacharjee1997distribution} to the simulation results. Here, $R$ is the head-hinge distance in the O shape and $L=51\nm$ is the length of the CC in the simulation. 
As we varied $\epsilon_b^{El}$, $l_p^{CC}$ changes, resulting in different $\Delta R_s$ values (Fig.~\ref{fig:dists_lp}). 
For $l_p^{CC} \approx 24 \nm $, we find that  $\Delta R_s=22\nm$, which coincides with the experimental value~\cite{ryu2020condensin}.  
The value of $l_p^{CC}$ is roughly six times larger than the experimental data. This may be due to the differences in the fitting used in the simulations or the errors in accurate measurements or in the method used to extract $l_p^{CC}$ from the experimental data. It should be emphasized that for comparison between theory and experiment $l_p^{CC}$ does not play an important rule. Nevertheless, it would be most interesting to design experiments to obtain precise estimates of $l_p^{CC}$.

The results in the inset in Fig.S3(a) are calculated using $l_p^{CC} \sim 4 \nm$. In this case, the elbow effect  is eliminated by setting $\epsilon^{CC}_b  = \epsilon^{El}_b =4 k_BT$, making  the entire CCs flexible. The simulations show that $\Delta R \approx 15\nm $, which disagrees with the experimental value ($\Delta R \approx 22\nm $).   We infer that  the existence of rigid portion of CC with flexibility in the elbow region  is essential for the scrunching mechanism.

\section{Derivation of $\boldsymbol{P(\vL|\vR)}$}
\begin{figure}[]
\centering
\includegraphics[width=0.5\textwidth]{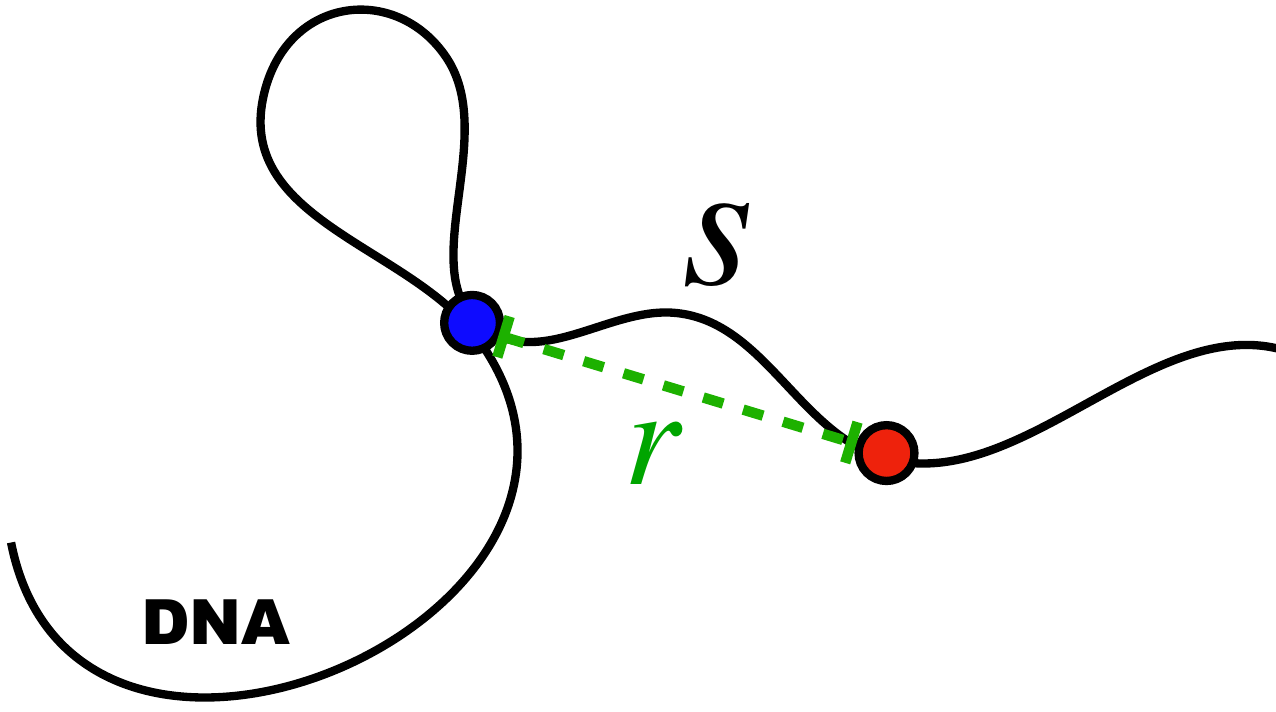}
\caption{\label{fig:derivation}A picture of a conformation of condensin bound to two loci separated by a genomic distance $s$ (extruded loop length). The  spatial distance between the attachment points in the DNA is $r$. For LE to occur condensin has to engage with at least two loci on the DNA.
}
\end{figure}
A major ingredient in the theory (see Eq.(1) in the main text) is the calculation of the contour length of the extruded loop as condensin is powered by ATP binding to the motor head, followed by hydrolysis, and subsequently resetting to complete the catalytic cycle. To obtain Eq.(1) in the main text, let us consider condensin separated by the spatial distance $r$ that pinches a loop whose genomic length is $s$ (Fig.\ref{fig:derivation}). Given the distribution of the spatial distance $r$ between two loci separated by a linear genomic distance $s$, $P(r|s)$, we would like to derive the distribution of $s$, $P(s|r)$. Indeed, $P(s|r)$ is the probability density of the extruded length of DNA, $s$, by condensin whose DNA binding domains are separated by the distance, $r$.  According to the Bayes theorem, we have $P(r|s)P(s) = P(s|r)P(r)$. The normalization dictates that $P(r) = \int_0^{L}P(r|s)P(s)\mathrm{d}s$. These two equations lead to,
\begin{align*}
P(s|r)=\frac{P(r|s)P(s)}{\int_0^{L}P(r|s)P(s)\mathrm{d}s}.
\end{align*}

We assume that there is no preference for picking a specific genomic distance $s$ on the DNA to which the motor attaches to initiate the extrusion process. Consequently,  we take $P(s)=1/L$. Therefore, 
\begin{align*}
P(s|r)&=\frac{(1/L)P(r|s)}{(1/L)\int_0^{L} P(r|s)\mathrm{d}s}\\
&=\frac{P(r|s)}{\int_0^{L}P(r|s)\mathrm{d}s}.
\end{align*}
Thus, $P(s|r)$ and $P(r|s)$ differ only by a constant, $\int_0^{L}P(r|s)\mathrm{d}s$, if we consider a fixed $r$.
It is clear that $P(r|s)$ is the radial probability density for the interior segments separated by a distance $r$ for a semi-flexible polymer, which is derived elsewhere \cite{hyeon2006kinetics}. For the case $s=L$, $P(r|s)$  is the result for the distribution of end-to-end distance for semi-flexible chains, $P(\vR|\vL)$~\cite{bhattacharjee1997distribution}. It is known that the simple analytic result for $P(\vR|\vL)$~\cite{bhattacharjee1997distribution} is accurate when compared to the exact result~\cite{Wilhelm96PRL} or numerical simulations. Thus, we employ the simpler expression $P(\vR|\vL)$ and assume that $P(\vL|\vR)$ is equivalent to $P(\vR|\vL)$ up to a normalization constant when expressed in terms of $L$ with fixed $R$.   Calculation of the distribution of loop sizes requires knowing, $P(\vL |\vR,f)$, which can also be derived from $P(\vR,f |\vL)$ in a similar manner.

\section{Effect of varying DNA persistence length}
\begin{figure}[]
\centering
\includegraphics[width=1.0\textwidth]{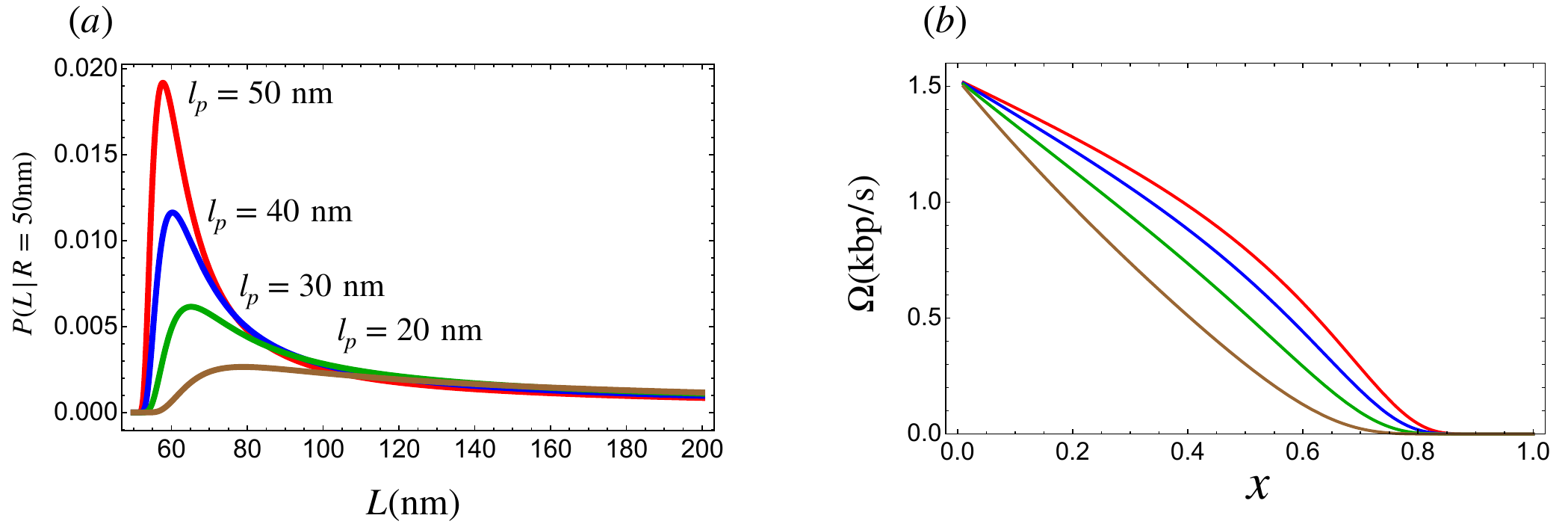}
\caption{\label{fig:rate_lp}Effect of variable persistence length for DNA. (a) Plot of $P(\vL|R=50\nm)$. $l_p=50\nm$ (red), $l_p=40\nm$ (blue), $l_p=30\nm$ (green), and $l_p=20\nm$ (brown). (b) Extrusion rate of DNA for different $l_p$. $l_p=50\nm$ (red), $l_p=40\nm$ (blue), $l_p=30\nm$ (green), and $l_p=20\nm$ (brown). $\DR$ is fixed to be $26\nm \sim 76\bps$. 
}
\end{figure}

In the main text, we used $l_p=50 \nm$ ($\sim$147 bps) as the persistence length of DNA, which is widely accepted value for DNA~\cite{rubinstein2003polymer}.
It could be interesting to explore the consequences of varying $l_p$, which can be drastically altered in the presence of divalent cations, as a variable in our theory. In Fig.\ref{fig:rate_lp}(a) we plotted $P(\vL|R=50\nm)$ using Eq.(1) in the main text for different $l_p$. As DNA becomes flexible the distribution of $P(\vL|R=50\nm)$ becomes wider, suggesting that most probable value of the captured length of DNA by condensin would be larger with a large dispersion. Thus, in this situation our approximation, $\vL \approx \vR$, would become less accurate. Nevertheless, we can explore the velocity of extrusion for different $l_p$ shown in Fig.\ref{fig:rate_lp}(b) for a fixed $\DR=26\nm$. As $l_p$ decreases, the velocity of extrusion becomes linear and slower because the load acting on DNA is higher for smaller $l_p$ at the same extension.The decrease of $\Omega$ as $l_p$ decreases can be deduced from the linear (small $x$) expansion of Eq.(7) in the main text. At small forces we find that  $f=\frac{3k_BT}{2 l_p}$. Substituting this linear expansion in the expression for $\Omega$ [Eq.(6)] confirms that as $l_p$ decreases $\Omega$ becomes smaller. Note  that this holds  for fixed extrusion length per step ($\sim$26 nm obtained by fitting Eq. (6) in the main text to the measured  LE velocity).

\section{Distribution of LE length per cycle}
\begin{figure}[]
\centering
\includegraphics[width=\textwidth]{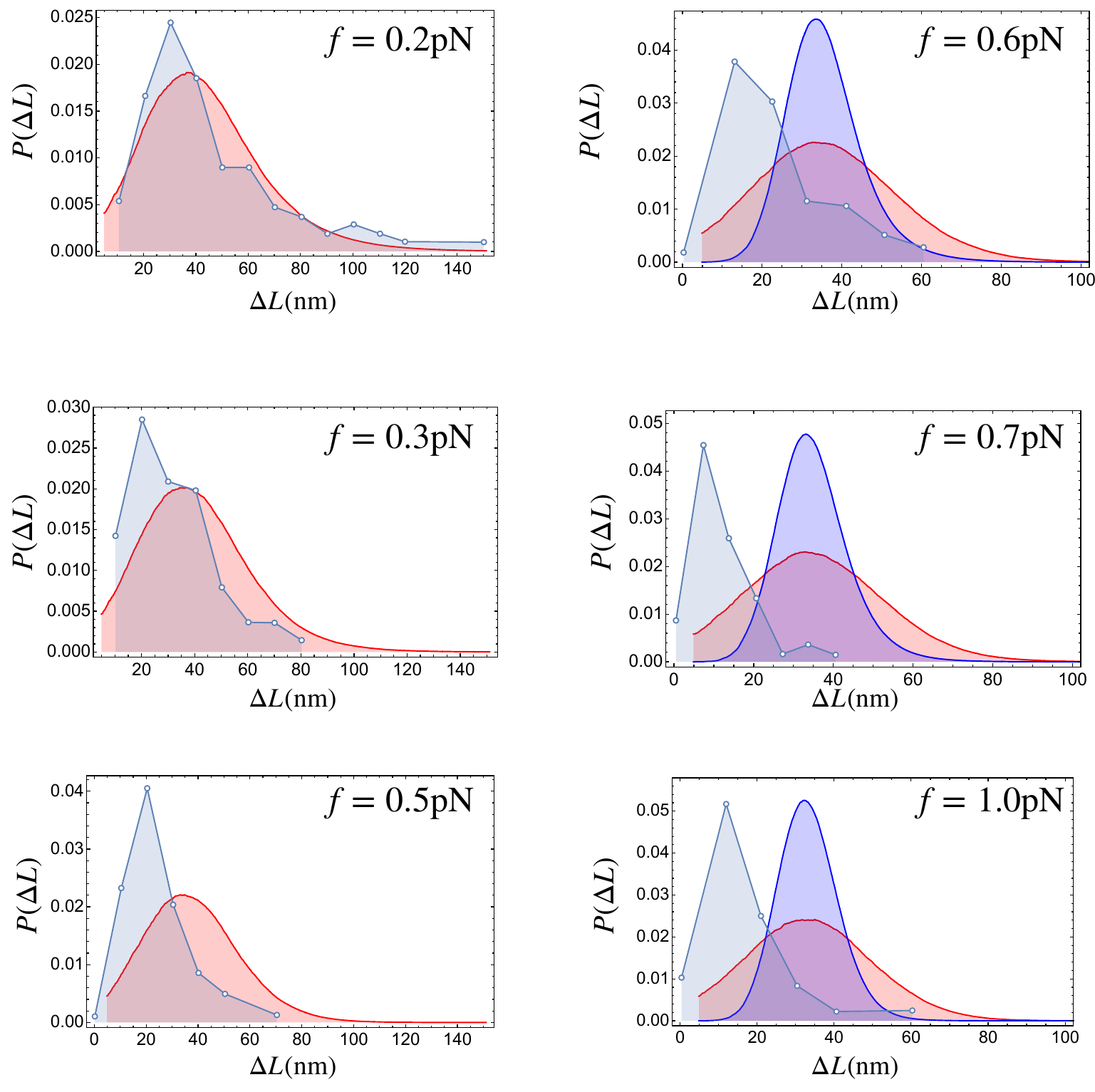}
\caption{\label{fig:dist_SI} Distributions for LE length per step for various external loads on DNA. The points in blue are from the  experiment~\cite{ryu2020resolving} and the distributions in red are from the theory (main text). The distributions in blue for high external loads are for the different standard deviation ($\Delta=5\nm$) for $R_1$.  
}
\end{figure}

In the main text, we showed that the theoretically derived distribution of LE length agrees well with the one obtained in the experiment~\cite{ryu2020resolving} for $f=0.4\pN$, {\it without adjusting any parameters}. The agreements persist for different values of load unless $0.5\pN \leq f$ (Fig.\ref{fig:dist_SI}). The discrepancies for high loads can be eased by using smaller $\Delta$ as shown in the blue distributions. It is worth mentioning that the sample sizes for obtaining the experimental results in Fig.~\ref{fig:dist_SI} are much smaller than for  $f=0.4\pN$.  They are $N=153,131,102,118,155,140$ for $f=0.2\pN,0.3\pN,0.5\pN,0.6\pN,0.7\pN,1.0\pN$ respectively whereas $N=1727$ for $f=0.4\pN$ in the main text. The smaller sample sizes surely affects the accuracy of the measured distribution. Considering the low stall load for condensin, likely to be $f \approx 0.8 \pN$ (see the inset of Fig.4a in the main text),  we believe that our theory predicts, with reasonable accuracy, the LE length by condensin during DNA compaction.

\section{Conformational transition to the LE active state}
\begin{figure}[]
\centering
\includegraphics[width=1.0\textwidth]{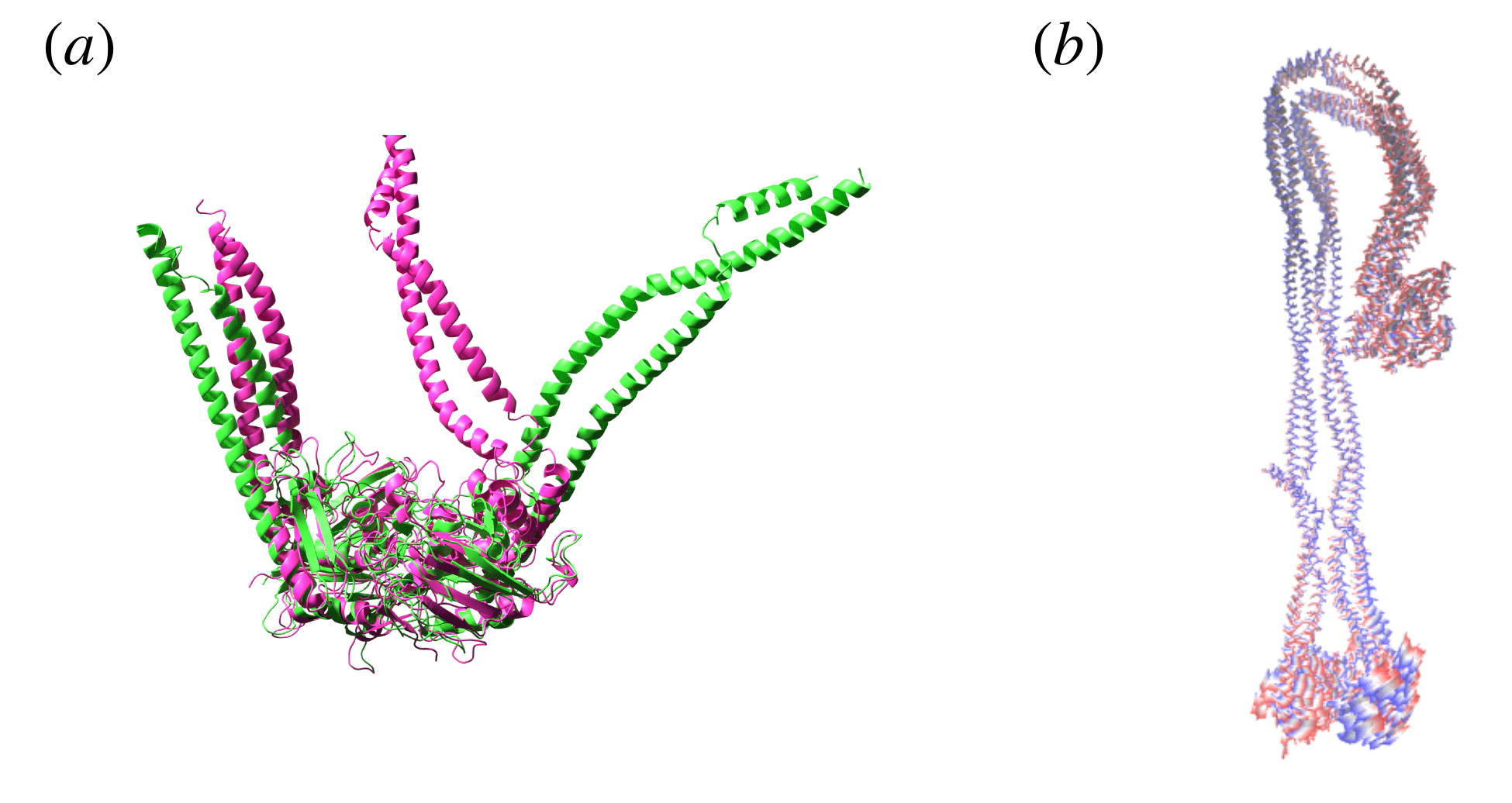}
\caption{\label{fig:alighn} (a) Structural alignment obtained by minimizing the root mean squared displacement for the head domain. Structure in green is the {\it apo} state (PDB:6YVU) where the heads are aligned to the ATP bound state, and the structure in magenta is the state with ATP (PDB:6YVD). 
Alignment is constructed only for the heads. (b) Normal mode analysis using the crystal structure for the {\it apo} state and ATP bound state (6YVU and 6YVD, respectively). Blurred regime shows the possible movements of the residues.  
}
\end{figure}
The lack of structures in various nucleotide states of condensin in the presence of DNA makes it virtually impossible to provide a molecular basis of loop extrusion. In the absence of viable structures, and prompted by the theoretical prediction that during a single turnover the distance between the hinge and motor should come within $\Delta R \approx (22-26) \nm$, we envisioned that the SMCs transition between the open and closed states in order to extrude loops. This picture is fully consistent with experiments~\cite{ryu2020condensin}. The shape of condensin from the partial structures of condensin in the inactive (in the absence of nucleotides or DNA) obtained by cryo-EM at $\sim 8.1 \angstrom$ resolution~\cite{lee2020cryo} cannot account for the O shape, which likely represents the functionally active state. Because the cryo-EM structures~\cite{lee2020cryo} do not contain DNA, they correspond to an inactive state.  Nevertheless, even these inactive structures~\cite{lee2020cryo} reveal that binding of ATP,  sandwiched between the two heads, leads to a substantial conformational change ($\sim 50 \degree$ rotation and $\sim 20 \angstrom$ translation) near the junction between the CC and the heads.

To provide insights into the ATP-induced conformational changes, we first aligned the structures in the head-domains of the {\it apo} state (PDB:6YVU) and the head-domains of the ATP-bound state (PDB:6YVD). Using VMD's multiseq tool~\cite{roberts2006multiseq}, we aligned residues: 1-148, 1036-1170 for Chain A and 150-311, 1285-1415 for Chain B using STAMP structural alignment.
The structural alignment shows that there is a large change in the  CC orientation between the {\it apo} state and the ATP bound state. 
Fig.~\ref{fig:alighn}a  shows the structural alignment obtained by minimizing the root mean squared displacement between the head domains in the {\it apo} state (PDB:6YVU), and the head domain in the ATP bound state (PDB:6YVD). The alignment suggests that the CC could undergo a wide opening motion upon ATP binding. 


{\bf Normal mode analysis:}
To understand the correlation between the head and hinge movement, we performed Normal Mode Analysis on the condensin structure in the {\it apo} state (PDB:6YVU). We created a variant of an Elastic Network Model using a second-order Taylor series expansion of the Self-organized Polymer model with Side-chains (SOP-SC) for proteins~\cite{mugnai2020role}. Since the {\it apo}-aligned to ATP bound state does not have the full condensin structure, we calculated a displacement vector ($\textbf{D}$) between 6YVU (aligned to 6YVD) with the original coordinate of 6YVU using only the beads that were present in both the full condensin structure, and the {\it apo}-aligned structure. Using the normal modes, we determined the eigenvector that best approximates this structural transition. $\textbf{D}$ is a $3N$ dimensional vector where N is the total number of beads  (PDB: 6YVD).  $\textbf{D}_i = [(\textbf{D})_{1,x},...,(\textbf{D})_{M,z}]$,
where $(\textbf{D})_{j,\alpha}$ is the entry associated with bead j, and direction $\alpha\in {x, y, z}$. The beads which are not present in the ATP-aligned structure have $(\textbf{D})_{j,\alpha}=0$.  The overlap between the eigenvector $v_n$ (corresponding to normal mode $n$), and displacement $\textbf{D}$ is given by,
\begin{equation}
    I_n(\textbf{D}) = \frac{\sum_{i=1}^{3N} v_{n,i}\cdot D_i }{\sqrt{\sum_{i=1}^{3N} v_{n,i}^2}\sqrt{\sum_{i=1}^{3N} \textbf{D}_i^2}}.
\end{equation}
We found that mode 7 has the highest overlap value.

Fig.~\ref{fig:alighn}b shows the motion of the residues calculated from the largest eigenvector that produces the maximum overlap. In the blurred region residues experience correlated movement. Our analysis shows that the normal modes mostly overlap at the head and hinge. In other words, head and hinge are likely to undergo the largest conformational change during the transition between the two states. If this finding holds when the structures of condensin in various nucleotide binding states are determined, it would provide a molecular basis for the O and B shapes that the condensin clearly samples during the process of loop extrusion~\cite{ryu2020condensin}. 

\bibliography{mybib}